\documentclass[a4paper,UKenglish,cleveref,autoref,thm-restate,authorcolumns]{lipics-v2019}
\usepackage{cite}
\usepackage[T1]{fontenc}
\usepackage{graphicx}
\usepackage[font={footnotesize}]{caption}
\usepackage[font={footnotesize}]{subcaption}
\usepackage{url}
\usepackage[normalem]{ulem}
\usepackage{amsmath}
\usepackage{mathtools}
\DeclarePairedDelimiter{\ceil}{\lceil}{\rceil}
\usepackage{tabularx}
\usepackage{listings}
\usepackage[capposition=top]{floatrow}
\usepackage{tabularx}
\usepackage{multirow}
\usepackage{threeparttable}
\usepackage{xcolor}
\usepackage{color}
\usepackage{flushend} % -- Added by RW to balance text on last page --
\usepackage{algorithm}
\usepackage{algpseudocode}

\newcommand{\annotatered}[1]{{#1}}

\algrenewcommand\algorithmicrequire{\textbf{Input:}}
\algrenewcommand\algorithmicensure{\textbf{Output:}}
\newcommand\ddfrac[2]{\frac{\displaystyle #1}{\displaystyle #2}}

\def\imagebox#1#2{\vtop to #1{\null\hbox{#2}\vfill}}

\definecolor{codegreen}{rgb}{0,0.6,0}
\definecolor{codegray}{rgb}{1,1,0.8}
\definecolor{codepurple}{rgb}{0.58,0,0.82}
\definecolor{backcolour}{rgb}{0.95,0.95,0.92}

\newcommand{\oursystem}{PAStime}
\newcommand{\litmusrt}{LITMUS\textsuperscript{RT}}

\def\BibTeX{{\rm B\kern-.05em{\sc i\kern-.025em b}\kern-.08em
		T\kern-.1667em\lower.7ex\hbox{E}\kern-.125emX}}

\nolinenumbers

\title{\oursystem{}: Progress-aware Scheduling for Time-critical Computing
}

\author{Soham Sinha}{Department of Computer Science, Boston 
University, USA}{soham1@bu.edu}{https://orcid.org/0000-0001-8962-4714}{}

\author{Richard West}{Department of Computer Science, Boston 
University, USA}{richwest@bu.edu}{https://orcid.org/0000-0001-5100-0666}{}

\author{Ahmad Golchin}{Department of Computer Science, Boston 
University, USA}{golchin@bu.edu}{}{}

\authorrunning{S. Sinha, R. West, and A. Golchin}

\Copyright{Soham Sinha, Richard West, and Ahmad Golchin}

\acknowledgements{This work is supported in part by the National Science 
Foundation
	(NSF) under Grant \#1527050. Special thanks to Dr. Ramesh Peri and
	Intel for generous support and feedback.  Any opinions, findings and
	conclusions or recommendations expressed in this material are those of
	the author(s) and do not necessarily reflect the views of the NSF.}

\EventEditors{Marcus V\"{o}lp}
\EventNoEds{1}
\EventLongTitle{32nd Euromicro Conference on Real-Time Systems (ECRTS 2020)}
\EventShortTitle{ECRTS 2020}
\EventAcronym{ECRTS}
\EventYear{2020}
\EventDate{July 7--10, 2020}
\EventLocation{Virtual Conference}
\EventLogo{}
\SeriesVolume{165}
\ArticleNo{6}
\begin{document}

\maketitle

\begin{abstract}%Real-time applications are normally configured to run with their worst-case
%parameters to guarantee their timely completion.  This leads to wastage of
%computing resources in an already resource-constrained environment, as
%worst-cases are practically rare.
%
Over-estimation of worst-case execution times (WCETs) of real-time tasks leads
to poor resource utilization. In a mixed-criticality system (MCS), the
over-provisioning of CPU time to accommodate the WCETs of highly critical
tasks may lead to degraded service for less critical tasks.  In this paper we
present \oursystem{}, a novel approach to monitor and adapt the runtime
progress of highly time-critical applications, to allow for improved service
to lower criticality tasks. In \oursystem{}, CPU time is allocated to 
time-critical tasks according to the delays they experience as they progress 
through their control flow graphs. This ensures that as much time as possible 
is made available to improve the Quality-of-Service of less critical tasks, 
while high-criticality tasks are compensated after their delays.

This paper describes the integration of \oursystem{} with Adaptive
Mixed-criticality (AMC) scheduling. The LO-mode budget of a
high-criticality task is adjusted according to the delay observed at
execution checkpoints. \annotatered{This is the first implementation of AMC in 
the scheduling framework of \litmusrt{}, which is extended with our 
\oursystem{} runtime policy and tested with real-time Linux 
applications such as object classification and detection.} We observe in 
our experimental evaluation that AMC-\oursystem{} significantly improves the 
utilization of the low-criticality tasks while guaranteeing service to 
high-criticality tasks.

\end{abstract}

%\begin{IEEEkeywords}
%	real-time systems, mixed-criticality systems, control flow graph, runtime system
%\end{IEEEkeywords}

\vspace{-6pt}
\section{Introduction}
\label{sec:introduction}
\vspace{-4pt}
In real-time systems, computing resources are typically allocated according to
each task's worst-case execution time (WCET), to ensure timing constraints are
met. However, worst-case conditions for an application are rather rare,
resulting in poor resource utilization.  Previous research
work~\cite{wilhelm2008worst} shows that the worst-cases lie at the tiny
tail-end of the probability distribution curve of the execution times for many
programs.  Instead, average-case execution times are significantly more
likely, taking a fraction of the worst-case times.  

Mixed-criticality systems (MCSs) provide a way to avoid
over-estimation of resource needs, by considering the schedulability
of tasks according to different estimates of their execution times at
different criticality or assurance
levels~\cite{vestal2007preemptive}. Higher criticality tasks are
afforded more execution time at the cost of less time for lower
criticality tasks, when it is impossible to meet all task timing
constraints. There have been multiple
proposals~\cite{de2009scheduling, baruah2008schedulability,
baruah2011response, burns2013mixed} since Vestal's work on
MCSs~\cite{vestal2007preemptive}.  Most prior work focuses on meeting
task deadlines and ignores other Quality-of-Service (QoS) metrics
\cite{vanga2017supporting} or average utilization.
Although timely completion is most important for high-criticality
applications, QoS is a significant metric for lower criticality tasks 
\cite{burns2013towards, su2013elastic, liu2016edf}.  This has motivated our 
work on \oursystem{} 
(\textbf{P}rogress-\textbf{A}ware \textbf{S}cheduling for \emph{time}-critical 
computing), to maximize the QoS for low-criticality
tasks.

In \oursystem{}, we first identify a checkpoint in a high-criticality
application at an intermediate stage in its source code.  The
application is then profiled offline to measure the time to reach the
marked checkpoint. Using this timing data, the application evaluates
its progress at the checkpoint during runtime. Based on the delay at
the checkpoint, \oursystem{} predicts the expected execution time of a
high-criticality application. We consequently adjust the runtime of
the application, given that the change does not hamper schedulability
of co-running tasks. If at runtime a highly critical program is
deemed to be making insufficient progress, it is given greater CPU
time.

We combine \oursystem{} with Adaptive Mixed-criticality (AMC)
scheduling~\cite{baruah2011response}, to improve the performance of
low-criticality tasks. In AMC, the system is started in {\em LO-mode},
where all tasks are scheduled with their LO-mode budgets. When a
high-criticality task runs for more than its LO-mode budget, the
system is switched to {\em HI-mode}. In HI-mode, all low-criticality
tasks are stopped, and the high-criticality tasks are given their
increased HI-mode budgets.  However, switching to HI-mode should be
avoided as much as possible
\cite{baruah2010towards}, because it affects the performance of low-criticality 
tasks, which are not executed in HI-mode.

Several works extend the mixed-criticality task model to improve the
performance of low-criticality tasks, such as providing an offline
extra budget allowance to the high-criticality
tasks~\cite{santy2012relaxing}, and estimating multiple
budgets~\cite{liu2018scheduling,ramanathan2018multi} and
periods~\cite{su2013elastic,su2016fixed} for low-criticality
tasks. However, these approaches do not utilize runtime information.

We extend AMC with \oursystem{} to avoid mode switches, by dynamically
adjusting the LO-mode budget for a high-criticality task, based on
progress to execution checkpoints. Then, we predict the expected
execution time of a high-criticality task based on the observed delay
until a checkpoint. We carry out an efficient online schedulability
test to determine whether we can increase the LO-mode budget of the
delayed high-criticality task to our predicted execution time. In case
the taskset is still schedulable with the increased budget, we extend
the LO-mode budget of the high-criticality task to the predicted
execution time. When a high-criticality task finishes within its
extended budget, we keep the system in LO-mode and avoid a
mode switch that would otherwise happen in AMC.
Thus, \oursystem{} improves the QoS of low-criticality tasks by
keeping the system in LO-mode for a longer time.

Factors such as I/O events and hardware microarchitectural delays
lead to actual execution times exceeding those predicted
by \oursystem{}. Any high criticality task running at the end of its
predicted execution time causes a timer interrupt to switch the system
into HI-mode, as is the case with AMC scheduling. Thus, a
high-criticality task never misses its deadline.

\vspace{-12pt}
\subsection{Contributions}
\vspace{-4pt}
The central idea of \oursystem{} is to help the OS make scheduling
decisions based on a program's runtime progress. 
\annotatered{This work is the first implementation of AMC in the scheduling 
framework of \litmusrt{} 
~\cite{calandrino2006litmus,brandenburg2011scheduling}. Such an implementation 
helps in testing AMC scheduling with a wide range of real-time Linux 
applications. We also implement an extension to AMC scheduling with our 
\oursystem{} runtime policy.} We test our implementation with real-world 
applications: an object classification application from the Darknet
neural network framework~\cite{darknet13}, an object tracking
application from the dlib machine learning toolkit~\cite{dlib}, and an
MPEG video decoder~\cite{ffmpeg18}.
We show that \oursystem{} increases the average utilization of
low-criticality tasks by 1.5 to 9 times for 2 to 20 tasks. We also
demonstrate that our implementation of AMC-\oursystem{} has minimal
and bounded additional overhead in \litmusrt{}.

We provide a C library for \oursystem{} to instrument checkpoints in
high-criticality programs. In addition, we modify the LLVM
compiler~\cite{lattner2004llvm} to \emph{automatically} identify potential
locations of checkpoints during profiling for simple time-critical
applications, \annotatered{and also instrument the checkpoints in the final 
binary executable file}.
	 
The next section describes our approach to \oursystem{}. 
Section~\ref{sec:theory} details the theoretical background behind AMC and its 
extension with \oursystem{}. Section~\ref{sec:design} describes the
design and implementation of \oursystem{}, which is then evaluated in
Section~\ref{sec:evaluation}.  Finally, we describe related work, followed by
conclusions and future work in Sections~\ref{sec:related}
and~\ref{sec:conclusion}, respectively.

\vspace{-6pt}
\section{Approach}
\label{sec:background}
\vspace{-8pt}
Compiler infrastructures such as LLVM are capable of producing a
program's control flow graph (CFG). A CFG represents the interconnection
between multiple {\em Basic Blocks (BBs)}, where a block is a sequence of
straight-line code without any internal branches.  However, CFGs are not
typically utilized by an OS to manage time for different 
computing resources, in spite of being a
rich source of analytical information about a program.  Consequently, current
OSs are oblivious to a program's computing requirements (e.g.,
CPU utilization) at different points in its execution. A developer of an 
application can, instead, help the OS make decisions related to resource 
management, by providing runtime information about a program at certain points 
in its source code.

\oursystem{} dynamically decides a program's execution budget based
on its runtime progress and theoretical analysis of the allowable
delay at specific checkpoint locations. At runtime, \oursystem{}
measures the time (i.e., CPU time) to reach a checkpoint from the start of a 
task, and then compares that time to a pre-profiled reference value. The
task's execution budget which was previously set based on profiling, 
is then adjusted according to actual progress.

\begin{figure}[t]
	\footnotesize
	\medskip
	\RawFloats
	\centering
	\caption{CFG and Average Time Estimates of a Program}
	\label{fig:simple_two_loop}
	\vspace{-6pt}
	\includegraphics[scale=0.8]{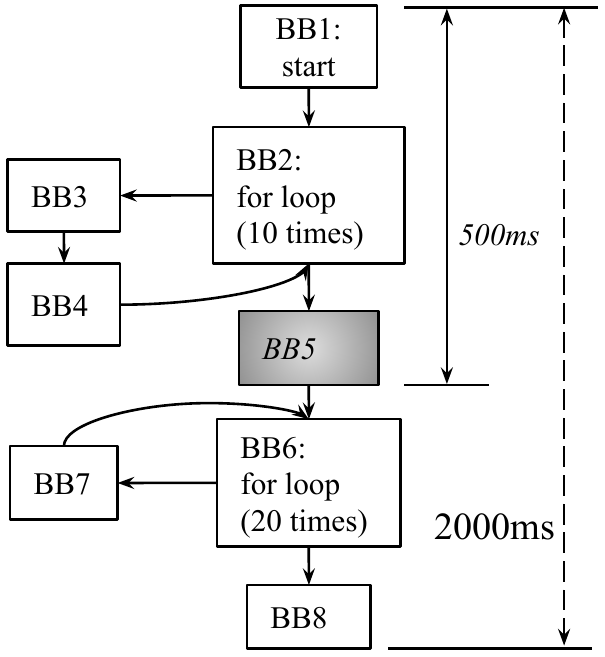}
	\vspace{-4pt}
\end{figure}

Figure~\ref{fig:simple_two_loop} shows the CFG for a program with two loops,
starting at BB2 and BB6.  
%Suppose that the average-case (AC) execution
%requirement of the program is a CPU utilization of 70\%. Here, utilization is
%defined as the percentage of time to the deadline that the program
%requires the CPU. 
In this example, \oursystem{} inserts a checkpoint between 
the two loops at
the end of BB5.  BB5 is a potentially good location for a
checkpoint because there is one loop before and after this BB, providing an
opportunity to increase the budget to compensate for the delay until 
BB5.

Suppose that we derive the LO-mode budget of the whole program to be
2000 ms by profiling, and the LO-mode time to reach the checkpoint at
BB5 is 500 ms.
%%
%% Soham, I just don't like the following text. It is too vague and weakens
%% the paper...
%% The BB5 checkpoint is assumed to be at a location in
%% the program where the difference between the task's LO- and HI-mode
%% progress is significant.  Otherwise, the LO-mode timing of 500 ms at
%% BB5 is ineffective in capturing a task's discrete LO-mode progress
%% from HI-mode progress.
%
%Given an average-case CPU utilization
%requirement of 70\%, offline profiling of the program determines the time to
%reach the checkpoint and also complete execution. 
%
%
The program is then executed in the presence of other tasks. The
execution budget at the checkpoint (in BB5) is adjusted, to account
for the program's actual runtime progress.  For example, suppose the
program experiences a delay of $100$ ms to reach the checkpoint in
BB5, thereby arriving at $600$ ms instead of the expected $500$
ms. Therefore, the program is delayed by ($\frac{100}{500}\times
100$=) 20\% from its LO-mode progress.

Depending on the relationship between the task's LO- and HI-mode
progress to the checkpoint, the task's budget is adjusted according to
the 20\% observed delay at the checkpoint. One approach is to
extrapolate a linear delay from the checkpoint to the end of the
task's execution.  Thus, the total execution time of 2000 ms is
predicted to complete at ($2000 + 20\% \times 2000$=) 2400 ms.
%This
%extrapolation is based on the runtime information until the checkpoint
%in BB5 which has shown a delay of 100 ms. Hence, we need to increase
%the program's execution budget by 400 ms.
\oursystem{} uses the
available information at a checkpoint to extend the LO-mode budget
of a high-criticality application. In
Section~\ref{sec:prediction_model}, we explore other possible
execution time prediction models.

%in LO-mode so that the schedulability of other tasks including the 
%low-criticality ones are not affected. We carry out an online 
%full schedulability test based on the offline calculated values 
%to determine if we can extend the task's LO-runtime by 400ms. We calculate a 
%safe upper bound on the duration of the online schedulability test. We explain 
%more about the online schedulability test in the next section.

\vspace{-4pt}
\subsection{Benefits of Adaptive Mixed-criticality Scheduling}
\vspace{-4pt}
We see progress-aware scheduling as being beneficial in
mixed-criticality systems. Adaptive Mixed-criticality
scheduling~\cite{baruah2011response} is the state-of-the-art
fixed-priority scheduling policy for Mixed-criticality tasksets. In
AMC, a system is first initialized to run in {\em LO-mode}. In
LO-mode, all the tasks are executed with their LO-mode execution
budgets.  Whenever a high-criticality task overruns its LO-mode
budget, the system is switched to {\em HI-mode}. In HI-mode, all
low-criticality tasks are discarded (or given reduced execution
time~\cite{liu2018scheduling,baruah2010towards}), and the
high-criticality tasks are given increased HI-mode budgets.  The
system's switch to HI-mode therefore impacts the QoS for
low-criticality tasks.

By combining \oursystem{} with AMC (to yield AMC-\oursystem{}), we
extend the LO-mode budget of a high-criticality task to its predicted
execution time at a checkpoint. Going back to our previous example in
Figure~\ref{fig:simple_two_loop}, we try to extend the LO-mode budget
of the task by 400 ms. We carry out an online schedulability
test to determine if we can extend the task's LO-mode budget by 400
ms. If the whole taskset is still schedulable after an extension of
the LO-mode budget of the delayed high-criticality task, the increment
in the task's LO-mode budget is approved. We let the task run until
the extended time and keep the system in LO-mode.  In case the
high-criticality task finishes within its extended time, we avoid an
unnecessary switch to HI-mode. Thus, the low-criticality tasks run for
an extended period of time and do not suffer from degraded CPU utilization
and QoS, as occurs with AMC.

In case the task does not finish within the predicted time, then the
system is changed to HI-mode, and low-criticality tasks are finally
dropped. This behavior is identical to AMC, and every high-criticality
task still finishes within its own deadline. Therefore, we improve the
QoS of the low-criticality tasks when the high-criticality tasks
finish within their extended LO-mode budgets, and otherwise, we fall
back to AMC.
%where we maintain at least the same level of service to
%the low-criticality tasks as is possible with AMC.
When there is no delay at a checkpoint, we do not change a task's LO-mode
budget.

%The idea is to improve service to low-criticality tasks while 
%trying to compensate the delays in high-criticality tasks. 
%%In this 
%%paper, we expect time-critical tasks are most important, which should not be 
%%delayed. 
%Low-criticality tasks gain improved 
%quality when given more CPU time, but do not have catastrophic consequences if 
%they are not serviced in a timely manner.
%In this paper, we propose an extension to the period transformation task model
%in mixed-criticality systems~\cite{vestal2007preemptive}, to provide a
%theoretical foundation for \oursystem{} in Section~\ref{sec:theory}. 
%\oursystem{} controls the CPU
%utilizations of the high-criticality (HC) tasks only, while attempting to
%improve the Quality-of-Service (QoS) of low-criticality (LC) tasks where 
%possible. 
%\vspace{-8pt}
%\subsubsection{Quality-of-Service of Low-criticality Task}
%\vspace{-4pt}

\vspace{-6pt}
\section{Theoretical Background}
\label{sec:theory}
\vspace{-6pt}
In this section, we provide a response time analysis for AMC-\oursystem{} by
extending the analysis for AMC scheduling.  We also describe details about the 
online schedulability test based on the response time values.

%\vspace{-14pt}
%\subsection{Schedulability Analysis for AMC and \oursystem{}}

\vspace{-4pt}
\subsection{Task Model}
\vspace{-4pt}
We use the same AMC task model as described by Baruah et
al~\cite{baruah2011response}.  Without loss of generality, we restrict
ourselves to two criticality levels - LO and HI. Each task, $\tau_i$, has five
parameters: $C_i(LO)$ - LO-mode runtime budget, $C_i(HI)$ - HI-mode runtime
budget, $D_i$ - deadline, $T_i$ - period, and $L_i$ - criticality level of a
task, which is either high ($HC$) or low ($LC$). We
assume each task's deadline, $D_i$, is equal to its period, $T_i$.  A HC
task has two budgets: $C_i(LO)$ for LO-mode assurance and
$C_i(HI)~(> C_i(LO))$ for HI-mode assurance.  A LC task has only
one budget of $C_i(LO)$ for LO-mode assurance.

\vspace{-4pt}
\subsection{Scheduling Policy}
\vspace{-4pt}
Both AMC and AMC-\oursystem{} use the same task priority ordering, based on
Audsley's priority assignment algorithm~\cite{audsley2001priority}. If a
task's response time for a given priority order is less than its period, then
the task is deemed schedulable. The details of the priority assignment
strategy are discussed in previous research
work~\cite{baruah2011response,baruah2013fixed}. We do not change the priority
ordering of the tasks at runtime.

AMC-\oursystem{} initializes a system in LO-mode with all tasks assigned their
LO-mode budgets. We extend a high-criticality task's LO-mode budget at a
checkpoint if the task is lagging behind its expected progress, as long as the
extension does not hamper the schedulability of the delayed task and all the
lower or equal priority tasks. An increase to the LO-mode budget of a task that
violates its own or other task schedulability is not allowed.  If a
high-criticality task has not finished its execution even after its extended
LO-mode budget, then the system is switched to HI-mode.

\vspace{-4pt}
\subsection{Response Time Analysis}
\vspace{-4pt}
The AMC response time recurrence equations for (1) all tasks in LO-mode, (2)
HC tasks in HI-mode, and (3) HC tasks during mode-switches are shown in
Equation~\ref{eq:LO_AMC},~\ref{eq:HC_HI_AMC} and~\ref{eq:HC_AMC_Star},
respectively. $hp(i)$ is the set of tasks with priorities higher than or equal
to that of $\tau_i$. Likewise, $hpHC(i)$ and $hpLC(i)$ are the set of high-
and low-criticality tasks, respectively, with priorities higher than or equal
to the priority of $\tau_i$.

AMC provides two analyses for mode switches: AMC-response-time-bound (AMC-rtb)
and AMC-maximum. We use AMC-rtb for our analysis, as AMC-maximum is
computationally more expensive. However, AMC-rtb does not allow a taskset
which is not schedulable by AMC-maximum. Therefore, AMC-rtb is {\em
sufficient} for schedulability.
\vspace{-4pt}
\begin{equation}
\begin{aligned}
R_i^{LO} = C_i (LO) + \sum_{\tau_j \in hp(i)} \ceil{\frac{R_i^{LO}}{T_j}} 
\times 
C_j (LO)
\label{eq:LO_AMC}
\end{aligned}
\end{equation}
\vspace{-16pt}
\begin{equation}
\begin{aligned}
R_i^{HI} = C_i (HI) + \sum_{\tau_j \in hpHC(i)} \ceil{\frac{R_i^{HI}}{T_j}} 
\times C_j (HI)
\label{eq:HC_HI_AMC}
\end{aligned}
\end{equation}
\vspace{-16pt}
\begin{equation}
\begin{aligned}
R_i^{*} = C_i (HI) &+ \sum_{\tau_j \in hpHC(i)} \ceil{\frac{R_i^{*}}{T_j}} 
\times C_j (HI)
&+ \sum_{\tau_j \in hpLC(i)} \ceil{\frac{R_i^{LO}}{T_j}} 
\times C_j (LO)
\label{eq:HC_AMC_Star}
\end{aligned}
\end{equation}
\vspace{-4pt}

%% --RW-- I don't see the following paragraph providing anything of
%% --RW-- significant value:
%% As the priority assignment strategy remains the same in AMC-\oursystem{} as
%% in the standalone AMC, the offline schedulability of random tasksets for
%% AMC-\oursystem{} remains also same. We dynamically extend the LO-mode
%% budgets of the delayed high-criticality tasks, only when the schedulability
%% of all the tasks with the offline priority ordering is not violated. We now
%% discuss the runtime schedulability issues with AMC-\oursystem{}.

With AMC-\oursystem{}, we not only measure $C_i(LO)$ and $C_i(HI)$,
but also the LO-mode time to reach a checkpoint $C_i^{CP}(LO)$. If a
high-criticality task, $\tau_i$, is delayed by $X\%$ at a checkpoint,
compared to its LO-mode progress, then $\tau_i$ takes
$(C_i^{CP}(LO) + \frac{C_i^{CP}(LO) \times X}{100})$ time to reach the
checkpoint. Hence, $\tau_i$'s budget is tentatively increased from
$C_i(LO)$ to $C_i^\prime(LO)$, where $C_i^\prime(LO) = f(C_i(LO),
X)$. Here, $f(C_i(LO), X)$ is a function to predict the delayed total
execution time, given that the observed delay until the checkpoint is
$X\%$. For example, by a linear extrapolation of the observed delay of
$X\%$ at a checkpoint, the original budget $C_i(LO)$ changes to
$f(C_i(LO), X) = \big(C_i(LO) + \ddfrac{C_i(LO) \times
  X}{100}\big)$. It is possible for $f$ to depend on task-specific and
hardware microarchitectural factors such as cache and main memory
accesses. We show more execution time prediction techniques in
Section~\ref{sec:prediction_model}.

With the increased LO-mode budget $C_i^\prime(LO)$, an online schedulability 
test then calculates the the extended
LO-mode response time, $R_i^{LO{\text{-ext}}}$, for $\tau_i$, using
Equation~\ref{eq:LO_AMC_Online}. Similarly, $R_i^{*{\text{-ext}}}$ is
calculated using
Equation~\ref{eq:HC_AMC_Star_Online}. Equations~\ref{eq:LO_AMC_Online}
and~\ref{eq:HC_AMC_Star_Online} are extensions of Equations~\ref{eq:LO_AMC}
and~\ref{eq:HC_AMC_Star}.  AMC-\oursystem{} checks at runtime whether both
$R_i^{LO{\text{-ext}}}$ and $R_i^{*{\text{-ext}}}$ are less than or equal to
$\tau_i$'s period to determine its schedulability.

The new response times are then calculated for all tasks in LO-mode with
priorities less than or equal to $\tau_i$, using
Equation~\ref{eq:LO_AMC_Online}. Similarly, new response times are calculated
for all HC tasks with lower or equal priority to $\tau_i$ during a mode
switch, using Equation~\ref{eq:HC_AMC_Star_Online}.  If all newly calculated
response times are less than or equal to the respective task periods, the
system is schedulable. In this case, AMC-\oursystem{} approves the LO-mode
budget extension to $\tau_i$. If the system is not schedulable, then
$\tau_i$'s budget remains $C_i(LO)$.

\vspace{-12pt}
\begin{equation}
\begin{aligned}
R_i^{LO{\text{-ext}}} = C_i^\prime (LO) &+ \sum_{\tau_j \in hp(i)} 
\ceil{\frac{R_i^{LO{\text{-ext}}}}{T_j}} 
\times C_j^\prime (LO)
\label{eq:LO_AMC_Online}
\end{aligned}
\end{equation}
\vspace{-10pt}
\begin{equation}
\begin{aligned}
R_i^{*{\text{-ext}}} = C_i (HI) &+ \sum_{\tau_j \in hpHC(i)} 
\ceil{\frac{R_i^{*{\text{-ext}}}}{T_j}} 
\times C_j (HI)
&+ \sum_{\tau_j \in hpLC(i)} \ceil{\frac{R_i^{LO{\text{-ext}}}}{T_j}} 
\times C_j (LO)
\label{eq:HC_AMC_Star_Online}
\end{aligned}
\end{equation}
\vspace{-10pt}
%\begin{equation}
%\begin{aligned}
%R_i^{*} = C_i (HI) &+ \sum_{\tau_j \in [hpHC(i) - k^{th}]} 
%\ceil{\frac{R_i^{*}}{T_j}} 
%\times C_j (HI) \\&+ \sum_{\tau_j \in hpLC(i)} \ceil{\frac{R_i^{LO}}{T_j}} 
%\times C_j (LO)
%\label{eq:AMC_Star_Online}
%\end{aligned}
%\end{equation}

AMC-\oursystem{} only extends the LO-mode budget of a delayed HC task for its
current job. When a new job for the same HC task is dispatched, it starts with
its original LO-mode budget. If another request for an extension in LO-mode
for the same task appears, AMC-\oursystem{} tests the schedulability with the
maximum among the requested extended budgets. The system keeps track of the
maximum extended budget for a task and uses that value for online
schedulability testing.  We explain the AMC-\oursystem{} scheduling scheme
with an example taskset in Table~\ref{tab:example_1}.

\vspace{-4pt}
\begin{table}[!ht]
	\footnotesize
	\centering
	\caption{A Mixed-criticality Taskset Example}
	\label{tab:example_1}
	\vspace{-6pt}
	\renewcommand{\arraystretch}{1.2}
	\begin{tabularx}{\columnwidth}{|X|X|X|X|X|X|X|X|}\hline
		\textbf{Task} & \textbf{Type} & \textbf{C(LO)} & \textbf{C(HI)} & 
		\textbf{T} & \textbf{Pr} &\textbf{$R^{LO}$} & $R^*$\\\hline 
		\hline 
		$\tau_1$ & HC & 3 & 6 & 10 & 1 & 3 & 6\\\hline
		$\tau_2$ & LC & 2 & - & 9 & 2 & 5 & - \\\hline
		$\tau_3$ & HC & 5 & 10 & 50 & 3 & 15 & 38\\\hline
	\end{tabularx}
	\vspace{-10pt}
\end{table}

Suppose, task $\tau_1$ is delayed by 66\% at a checkpoint in the
task's source code. \oursystem{} will then try to extend the budget by
$(3\times\frac{66}{100})\approx 2$ time units.  Therefore, the
potential extended budget $C_1^\prime(LO)$ for $\tau_1$ would be
$(3+2)=5$ time units. \oursystem{} will calculate the response times,
$R_i^{LO\text{-ext}}$ and $R_i^{*\text{-ext}}$, for $\tau_1$ and the
lower priority tasks $\tau_2$ and $\tau_3$. $R_1^{LO{\text{-ext}}}$
would just be 5, and $R_1^{*{\text{-ext}}}$ would remain the same as
$R_1^*=6$, as $\tau_1$ is the highest priority task.  The new
$R_2^{LO{\text{-ext}}}$ would be 7 (by
Equation~\ref{eq:LO_AMC_Online}) which is smaller than its period of
9. Therefore, $\tau_2$ would still be schedulable if we extend
$\tau_1$'s LO-mode budget from 3 to 5.

For $\tau_3$, the new $R_3^{LO{\text{-ext}}}$ and
$R_3^{*{\text{-ext}}}$ would be, respectively, 26 (by
Equation~\ref{eq:LO_AMC_Online}) and 40 (by
Equation~\ref{eq:HC_AMC_Star_Online}) which are also smaller than
$\tau_3$'s period of 50. Therefore, the extended budget of $5$ for
$\tau_1$ would be approved by AMC-\oursystem{}. In conventional AMC
scheduling, the system would be switched to HI-mode if $\tau_1$ did
not finish within 3 time units. However, AMC-\oursystem{} will extend
$\tau_1$'s LO-mode budget to 5 because of the observed delay at its
checkpoint, so the system is kept in LO-mode. Consequently, LC task
$\tau_2$ is allowed to run by AMC-\oursystem{}, if $\tau_1$ finishes
before 5 time units.  If $\tau_1$ does not finish even after 5
time units, the system would be switched to HI-mode.

In this example, 5 jobs of $\tau_1$ are dispatched for every single
job of $\tau_3$, as $\tau_3$'s period of 50 is 5 times the period of
$\tau_1$.  Suppose $\tau_1$ asks for the 66\% increment in its LO-mode
budget for the first job, as we have explained above. In its second
job, $\tau_1$ asks for an increment of 33\% (1 time unit) in its
LO-mode budget.  In this case, we again need to calculate the response
times for all tasks.  If we calculate the online response times
for $\tau_2$ and $\tau_3$ assuming $(3+1)=4$ time units for
$C_1^\prime(LO)$, then we would not account for the first job of
$\tau_1$, which potentially executes for 5 time units. We would need
to keep track of the extended LO-mode budgets for all jobs of $\tau_1$,
to accurately calculate online response times for $\tau_2$ and
$\tau_3$.

To avoid the cost of recording all extended LO-mode budgets for a
task, AMC-\oursystem{} simply stores the maximum extended budget for a
task.  When calculating online response times to check whether to
approve an extension to the LO-mode budget, the system uses the
maximum extended budget of every high-criticality task. This value is
stored in the \texttt{max\_extended\_budget} variable for each HC
task.  Therefore, when $\tau_1$ asks for $4$ time units as its
extended LO-mode budget in its second job, the system calculates
$R_1^{LO{\text{-ext}}}$, $R_1^{*{\text{-ext}}}$,
$R_2^{LO{\text{-ext}}}$, $R_3^{LO{\text{-ext}}}$,
$R_3^{*{\text{-ext}}}$ with $C_1^\prime(LO)=5$.

AMC-\oursystem{} uses maximum extended budgets to calculate safe upper
bounds for online response times. The \texttt{max\_extended\_budget} task 
property is reset when a task has not requested a LO-mode budget extension for
any of its dispatched jobs within the maximum period of all tasks.

\vspace{-12pt}
\subsection{Online Schedulability Test}
\vspace{-4pt}
AMC-\oursystem{} performs an online schedulability test whenever
a high-criticality task asks for an extension to its LO-mode
budget. The test calculates the response times ($R_i^{LO\text{-ext}}$
and $R_i^{*\text{-ext}}$) of the delayed high-criticality task and all
lower priority tasks. Then, it checks whether the response times are
less than or equal to the task periods. If any task's response time is
greater than its period, the online schedulability test returns false,
and the extension in LO-mode for the high-criticality task is
denied. If the schedulability test is successful the high-criticality
task is permitted to run for its extended budget in LO-mode.

%\vspace{-12pt}
%\subsubsection{Details of the Online Schedulability Test}
%\vspace{-4pt}
\begin{algorithm}[t]
	\caption{Online Schedulability Test}
	\label{algo:change_budget_runtime}
	%\hspace*{\algorithmicindent} \textbf{Input:}  All\_Tasks\\
	%\hspace*{\algorithmicindent} \textbf{Output:} \\ 
	\begin{algorithmic}[1]
		\State \textbf{Input:} $tasks$ - set of all tasks in priority order
		\State        ~~~~~~~~~$\tau_k$ - delayed task
		\State        ~~~~~~~~~$e$ - extra budget for $\tau_k$
		\State \textbf{Output:} {$true$ or $false$}
		\Function{IsBudgetChangeApproved}{$tasks$, $\tau_k$, $e$}
		\State $e^\prime$ = max$(\tau_k.$max\_extended\_budget$, C_k(LO)+e)- 
		C_k(LO)$ 
		\label{lst:line:new_e_max}
%		\State ~~~~~~
		\For{each task $\tau_i$ in \{$\tau_k$ $\cup$ lower priority tasks than 
			$\tau_k$ in $tasks$\}}
		\State $C_i^\prime(LO) = \tau_i.$max\_extended\_budget 
		\label{lst:line:max_budget_start}
		\If{$\tau_i$ is $\tau_k$}
		\State $C_i^\prime(LO) = $max$(C_i^\prime(LO), C_i(LO) + e)$ 
		\label{lst:line:c_i_dash_low}
		\EndIf \label{lst:line:max_budget_end}
		\State Initialize $R_i^{LO{\text{-ext}}}$ for 
		Equation~4 with 
		$R_i^{LO} + e^\prime$
		\State Solve Equation~4 
		\label{lst:line:eq_lo_amc_online}
		\If{$R_i^{LO{\text{-ext}}}$ $\leq T_i$}
		\If{$\tau_i$ is high-criticality}
		\State Initialize $R_i^{*{\text{-ext}}}$ for 
		Equation~4 with $R_i^{*}$
		\State Solve Equation~5 with the new
		$R_i^{LO{\text{-ext}}}$ \label{lst:line:solve_hc_amc}
		\If{$R_i^{*\text{-ext}}$ $> T_i$} \textbf{ Return} $false$
		\EndIf
		\EndIf
		\Else \textbf{ Return} $false$
		\EndIf
		\EndFor
		\State $\tau_k.$max\_extended\_budget = max($C_k(LO) + e, 
		\tau_k.$max\_extended\_budget)
		\State \textbf{Return} $true$
		\EndFunction
	\end{algorithmic}
      \end{algorithm}

Algorithm~\ref{algo:change_budget_runtime} shows the pseudocode for the
online schedulability test. The offline response time values
($R_i^{LO}$ and $R_i^{*}$) are stored in the properties for
each task $\tau_i$. When a high-criticality task $\tau_k$ is delayed,
it asks for an extension of its LO-mode budget by $e$ time units.

Line~\ref{lst:line:new_e_max} in Algorithm~\ref{algo:change_budget_runtime} 
determines whether the newly requested extension, $e$, is more than a previously
saved maximum extended budget for $\tau_k$.  In 
lines~\ref{lst:line:max_budget_start}--\ref{lst:line:max_budget_end}, the
$C_i^\prime(LO)$ is set to the maximum extended budget for all the lower 
priority tasks and $\tau_k$. As we have
explained above, it is practically infeasible to store the execution
times of every job of all the tasks to calculate online response
times. Therefore, we store the maximum extended budget of every task
and use this to calculate response times online. For the currently
delayed task, $\tau_k$, $C_k^\prime(LO)$ is set to the maximum value
between a previously saved \texttt{max\_extended\_budget} and the
currently requested $C_k(LO) + e$. Considering the example
in Table~\ref{tab:example_1} from the paper, line~\ref{lst:line:c_i_dash_low} would translate 
to $C_i^\prime(LO)$=max(5, 4) for the second LO-mode extension request.

Then, Equation~4 is solved in 
line~\ref{lst:line:eq_lo_amc_online} by
initializing $R_i^{LO\text{-ext}}$ to the $R_i^{LO}$ plus extra budget
$e^\prime$ from line~\ref{lst:line:new_e_max}. If a lower priority task is a 
high-criticality task, then
$R_i^{*\text{-ext}}$ is calculated in line~\ref{lst:line:solve_hc_amc}, with 
the newly derived value of $R_i^{LO\text{-ext}}$. 

%The initial value in the recurrence relations is set {\em offline} to
%significantly decrease the number of {\em online} iterations, and
%hence overhead, needed to determine taskset schedulability.  In
%addition, we solve both recurrence relations with an incremental
%response time algorithm to further minimize the overhead.

% above lines are only needed if the next subsubsection is not added.

%\vspace{-10pt}
%\subsubsection{Initial Value for Online Response Times Calculation}
%\vspace{-4pt}
Online response time calculations may take significant time, depending
on the number of iterations of Equations~\ref{eq:LO_AMC_Online}
and~\ref{eq:HC_AMC_Star_Online}. However, the response time values
from
Equations~\ref{eq:LO_AMC},~\ref{eq:HC_HI_AMC},~\ref{eq:HC_AMC_Star}
are already calculated offline to determine the schedulability of a
taskset using Audsley's priority assignment
algorithm \cite{audsley2001priority} with AMC scheduling. Since
AMC-\oursystem{} uses the same priority ordering as AMC,
the offline response times for schedulability remain the same.

AMC-\oursystem{} initializes the online $R_i^{LO\text{-ext}}$ in
Equation~\ref{eq:LO_AMC_Online} with $R_i^{LO} + e$, where $R_i^{LO}$
is calculated offline by Equation~\ref{eq:LO_AMC} and $e(>0)$ is the
extra budget of the delayed task.

Since the budget of a delayed task is extended by $e$,
$R_i^{LO\text{-ext}}$ must be greater than or equal to $R_i^{LO} + e$.
Hence, this is a good initial value to start calculating
$R_i^{LO\text{-ext}}$ online.  In Section~\ref{sec:evaluation}, we
establish an upper bound on the total number of iterations needed to
check the schedulability of random tasksets, if the highest priority
task's budget is extended by different amounts.

%Additionally, we also store the variable \texttt{max\_extended\_budget} for 
%a task if a high-criticality had been allowed a LO-mode extension in its 
%budget. We use these extended budgets when we recalculate the response times.

%\vspace{4pt}
%\begin{algorithm}
%\caption{Algorithm to change budget at runtime}
%\label{algo:change_budget_runtime}
%\begin{algorithmic}[1]	
%	\Function{Change\_Budget}{task, new\_budget}
%	\State prev\_budget = task.budget
%	\State diff = prev\_budget - new\_budget
%		\If{diff $\geq 0$ and Is\_Lower\_Priority\_Sched(task, task, diff)}
%			\textbf{ return} $1$
%		\Else
%			\textbf{ return} $0$
%		\EndIf
%	\EndFunction
%	\newline
%	\Function{Is\_Lower\_Priority\_Sched} {task, changed\_task, diff}
%		\State Get new LO\_Response\_Time of $task$ with diff in budget for 
%		$changed\_task$
%		\If{LO\_Response\_Time < period of $task$}
%			\If{task is HC}
%				\State Get new HI\_Response\_Time of $task$ with new 
%				LO\_Response\_Time
%				\If{HI\_Response\_Time < period of $task$}
%					\If{no lower priority task}
%						\textbf{ return 1}
%					\EndIf
%					\State Is\_Lower\_Priority\_Sched(next lower priority task, 
%					changed\_task, diff)
%				\Else
%					\textbf{ return 0}
%				\EndIf
%			\Else
%				\If{no lower priority task}
%					\textbf{ return 1}
%				\EndIf
%				\State Is\_Lower\_Priority\_Sched(next lower priority task, 
%				changed\_task, diff)
%			\EndIf
%		\Else
%			\textbf{ return 0}
%		\EndIf
%	\EndFunction
%\end{algorithmic}
%\end{algorithm}

\vspace{-6pt}
\section{Design and Implementation of \oursystem{}}
\label{sec:design}
\vspace{-4pt}
In this section, we first describe the overall design of
AMC-\oursystem{} in \litmusrt{} 
\cite{calandrino2006litmus,brandenburg2011scheduling}.
This is followed by a description of how checkpoints are instrumented
in an application's source code. We then show the algorithm to
determine and insert checkpoints, which is integrated into the LLVM
compiler infrastructure. A high-criticality task requires profiling to
determine the placement of checkpoints, before it is ready for
execution with other tasks. We describe how the profiling and
execution of a high-criticality task is performed, along with the
scheduling mechanism in \oursystem{}. The source code for \oursystem{}
in \litmusrt{} is publicly available \cite{pastime_code}.

AMC-\oursystem{} has two phases: a Profiling phase for
high-criticality tasks, and an Execution phase.  In the Profiling
phase, one or more checkpoints are placed at key stages in a program's
source code.  The {\em average} time to reach each checkpoint is then
measured. After profiling all high-criticality tasks, the system
switches into the Execution phase. The time taken to reach each
checkpoint in every high-criticality task is observed by the system at
runtime. Each observed time is compared against the profiled time to
reach the same checkpoint. Any high-criticality task lagging behind
its profiled time to a checkpoint is tentatively given increased
LO-mode budget, according to the approach described in
Section~\ref{sec:theory}.

\begin{figure}[t]
	\centering
	\caption{Implementation of AMC-\oursystem{} in \litmusrt{}}
	\label{fig:pastime-execution-flow}
	\vspace{-6pt}
	\includegraphics[scale=0.6]{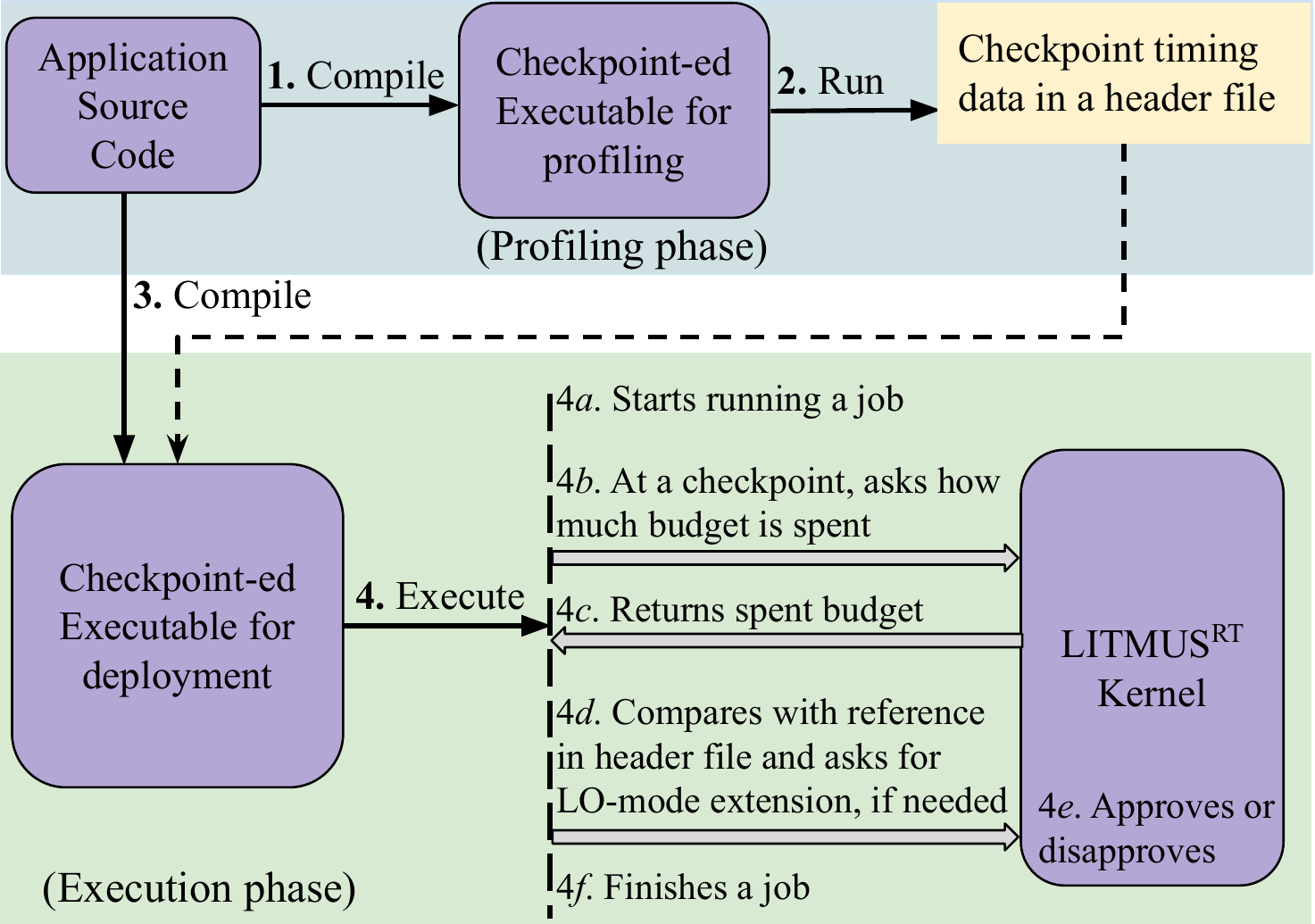}
	\vspace{-4pt}
\end{figure}

Figure~\ref{fig:pastime-execution-flow} shows an overview of the
design of AMC-\oursystem{} in \litmusrt{}.  Step 1 is the compilation
of a high-criticality application's source code in the Profiling
phase, which uses our compilation procedure~\cite{pastime_code}.  The
compiled executable has checkpoints embedded into its code for
profiling.  Step 2 executes the program to generate timing metadata
for each checkpoint in a \textit{timeinfo.h} file. Step 2 is performed
multiple times with different program inputs to generate an average
time to reach each checkpoint.

Step 3 compiles the source code along with the checkpoint timing
metadata header file (\textit{timeinfo.h}) for the Execution phase,
producing a binary image that is used for deployment under working
conditions. Finally, Step 4 runs the code in the Execution phase along
with all other tasks.  At some point after the system is started, Step
4a starts running a job for a high-criticality task.  When a
checkpoint is reached, Step 4b asks the \litmusrt{} kernel how much
budget it has consumed. Step 4c returns the spent budget from
the \litmusrt{} kernel to the application.

After receiving the spent budget, $t_{\text{spent}}$, the application
compares it to the reference timing, $t_{\text{ref}}$, for the
checkpoint from the \textit{timeinfo.h} header file. It calculates the
extra budget, $e$, needed in LO-mode, using an execution time
prediction model as described later in
Section~\ref{sec:prediction_model}. If $e>0$, Step 4d asks \litmusrt{}
for extra budget. The AMC-\oursystem{} scheduling policy in
the \litmusrt{} kernel runs an online schedulability test. If the test
returns true, the LO-mode budget of the current task is extended by
$e$. If the test algorithm returns false, the task budget is not
altered. Finally, Step 4f finishes the current job of the running
task.

%\vspace{-6pt}
%\begin{equation}
%\begin{aligned}
%e = \ddfrac{C_i(LO) \times {(t_\text{spent} - t_\text{ref})}}{t_\text{ref}}
%\label{eq:extra_budget}
%\end{aligned}
%\end{equation}

\vspace{-8pt}
\subsection {Checkpoint Instrumentation and Detection}
\label{subsec:checkpoint}
\vspace{-4pt}
A checkpoint is a key stage in an application's code, used to evaluate
the progress of a currently running task. Well placed checkpoints
balance the number of instructions that are executed prior to the
checkpoint, with those that remain to the next checkpoint or the end
of the program. Ideally, there should be a meaningful number of
instructions leading up to a checkpoint to determine
progress. Likewise, there should be sufficient instructions after a
checkpoint to increase the likelihood that a task is adequately
compensated for execution delays using an extended budget.
%Here, we explain the strategy to decide the location of a checkpoint.
%A checkpoint is instrumented in a program's source code for both 
%the Profiling and Execution phase.

A developer of a high-criticality application finds a potential
checkpoint location in the program's source code for its Execution
phase, after trying out multiple locations in the Profiling
phase. \oursystem{} includes a development library to instrument
checkpoints for the two different phases. Additional modifications to
the LLVM compiler~\cite{lattner2004llvm} are used to automatically detect and
instrument checkpoints in the Profiling phase.

%Although each 
%instruction may exhibit timing
%variation, loops are potential portions of the code where greater execution
%time variation occurs. 
%Therefore, we adopt a policy \emph{to insert
%checkpoints between two consecutive loops}.

\vspace{-8pt}
\subsection{Checkpoint Library}
\label{subsec:user-space-library}
\vspace{-4pt}
We have developed a C library to instrument checkpoints in the
two \oursystem{} phases.  The main purpose of the library is to
generate the necessary checkpoint timing information during the
Profiling phase and then request a task's extended LO-mode budget from
the \litmusrt{} kernel during the Execution phase.

In the Profiling phase, a \texttt{writeTime} function call from our
library is inserted into the application code at a desired
checkpoint. \texttt{writeTime} takes a unique ID for each checkpoint. The
function logs the time to reach that checkpoint in the source code
since the start of a job.  The average of multiple such timing entries
is saved in \textit{timeinfo.h} after the Profiling phase.

During the Execution phase, a preprocessor macro
called \texttt{ANNOUNCE\_TIME} is inserted at a checkpoint. This macro
obtains the spent budget from the \litmusrt{} kernel via
a \texttt{get\_current\_budget} system call. Then, it compares the
spent budget with the reference budget from the \textit{timeinfo.h}
header file, and calculates the extra budget using
an execution time prediction model.  If extra budget is needed, the macro
makes a \texttt{set\_rt\_task\_param} \litmusrt{} system call.

\vspace{-4pt}
\subsection{Manual Checkpoint Instrumentation}
\vspace{-4pt}
A developer may attempt various strategies to identify a key
stage~\cite{burr1998combinatorial, tikir2002efficient,trucov} of an
application to instrument a checkpoint. The developer uses either
the \texttt{writeTime} function for the Profiling phase, or
the \texttt{ANNOUNCE\_TIME} macro for the Execution phase to
instrument a checkpoint. Checkpoints should generally be avoided
inside tight loops. Visiting a checkpoint every loop iteration incurs
a small overhead that is accumulated across multiple iterations.

\vspace{-4pt}
\subsection{Automatic Checkpoint Instrumentation}
\label{sec:auto_checkpoint}
\vspace{-4pt}
We have written a compiler pass in LLVM to automatically instrument
checkpoints for the Profiling phase of \oursystem{}. The instrumented
code is run in the Profiling phase, and multiple checkpoint timing
information is generated.  Finally, the developer chooses one such
checkpoint for the Execution phase.

The compiler pass automatically inserts checkpoints in the basic block
preceding each loop in a function, except the first loop. The first
loop is excluded so that enough instructions are executed before a
checkpoint to determine meaningful delays.

For nested loops, we consider only the outer loop. Automatic
instrumentation works with only simple program structures and ignores
intersecting loops. LLVM's
\texttt{LoopInfo} analysis identifies only natural loops~\cite{loopinfo}. 
We utilize the \texttt{LoopInfo} class in our checkpoint
instrumentation implementation.

\begin{algorithm}[t]
	\caption{Determine and Insert Checkpoints}
	\label{algo:checkpoint}
	\begin{algorithmic}[1]	
		\State $isLoopBefore$: identifies if a loop is in the paths from the starting BB to another BB
%		\State $Checkpoints$: tells if a Checkpoint is added 
%		for a $LoopID$
		\State $visited$: set of already visited BBs in DFS
		\Function{insertCheckpoint}{function}
		\State $startingBB$ = $function.getEntryBlock()$
		\State \Call{doDFS}{$startingBB$}
		\EndFunction
		\Function{doDFS}{$currentBB$}
		\If{$currentBB$ is in $visited$}
		\textbf{ return}
		\EndIf
		\State $LoopID$ = getLoopFor($currentBB$)
		\If{$LoopID$ != null}
		\If{$isLoopBefore[currentBB]$ $\land$ isLoopHeader($currentBB$)}
%			$LoopID$ not in $Checkpoints$}
		\State insert checkpoint before $currentBB$
%		\State add $LoopID$ in $Checkpoints$
		\EndIf
		\EndIf
		\State insert $currentBB$ in $visited$
		\For{each $s$ in successors of $currentBB$} \Call{doDFS}{$s$}
		\EndFor
		\EndFunction
	\end{algorithmic}
	\vspace{-4pt}
\end{algorithm}
%\vspace{-10pt}

Algorithm~\ref{algo:checkpoint} identifies checkpoint locations for
the Profiling phase. It starts a Depth First Search (DFS) from the
starting Basic Block (BB) of a function, by calling \texttt{DoDFS} in
line 6.  The algorithm then checks whether a BB is part of a loop
using \texttt{LoopInfo}'s \texttt{getLoopFor} member function. This
function returns a unique \texttt{LoopID} for every new loop. A
nested inner loop has the same ID as its outer loop. If a BB is not
part of a loop, then it returns \texttt{null}.

If a BB is part of a loop, line 12 first checks whether there is
any loop before the current loop using a
dictionary \texttt{isLoopBefore}.
\texttt{isLoopBefore} 
is pre-populated for every BB in the CFG to indicate whether there is
at least one loop seen in the paths from the starting BB to the
current BB. To pre-populate \texttt{isLoopBefore}, the algorithm checks
whether there is a path to a BB from the loop BBs.

Algorithm~\ref{algo:checkpoint} then verifies that the current
BB is a header of a loop
using \texttt{LoopInfo}'s \texttt{isLoopHeader} member function. A
header is the entry-point of a natural loop. We insert a checkpoint in
the predecessor BB of a header.
%\vspace{-1.9pt}

Finally, if there is at least one loop before the current header BB,
a checkpoint is added before the current BB in line 13. As
natural loops have only one header, a checkpoint is avoided
for any inner loop within a nested loop.
%Then, we add the $LoopID$ in \texttt{Checkpoints}. 
%As LLVM 
%returns same loop id for every BB part of a loop including the nested loops, 
%we 
%avoid inserting any checkpoint for any inner loop by checking the 
%\texttt{Checkpoints}. 
We continue the DFS by marking the current BB as visited.

We have implemented the above algorithm in a compiler pass within
LLVM~\cite{lattner2004llvm}, to automatically detect and instrument
appropriate function calls as checkpoints in a C language
program. This uses our previously described Checkpoint library for the
Profiling phase.  A developer uses our modified LLVM compiler with
their high-criticality application written in C.  Our compiler pass
operates at the LLVM Intermediate Representation (IR) level.  It takes
a piece of IR logic as input, figures out the points of interest
according to the above algorithm for checkpoints in a particular
function, and generates the instrumented IR.  These IRs are compiled
into executable machine code.  As the compiler pass operates at the IR
level, it is easily extensible to other high-level languages and
back-ends supported by LLVM.

%%  --RW-- I greatly shortened this section as most of it has already been
%% said earlier.
\vspace{-4pt}
\subsection{Profiling and Execution Phases}
\label{sec:phases}
\vspace{-4pt}
The Profiling phase of \oursystem{} determines viable checkpoints for
use in the Execution phase, and also the LO- and HI-mode budgets for a
high-criticality application. A checkpointed program is run multiple
times in the Profiling phase against a set of test cases. The program is
allowed to have multiple test checkpoints, which are either generated
automatically using our modified LLVM compiler, or manually by a
developer. Each checkpoint has a unique ID (function ID, BasicBlock ID) given 
to the checkpoint function call \texttt{writeTime}. A Python 
package then runs the Profiling phase to collect the average times (LO-mode) to 
reach different checkpoints.

Step 4 in Figure~\ref{fig:pastime-execution-flow} is the start of the
Execution phase. One checkpoint from those generated in the Profiling phase
for a high-criticality application is instrumented with
an \texttt{ANNOUNCE\_TIME} macro. Although in general it is possible to use
multiple checkpoints within the same application in the Execution phase, our
experience shows that one is sufficient to improve LO-mode service. Multiple
checkpoints add overhead to the task execution. Moreover, later checkpoints
account for execution delays that make earlier checkpoints redundant, as long
as they are reached before the LO-mode budget expires.

The key issue in deciding on a single checkpoint for the Execution phase is to
ensure it is not placed too late in the instruction stream. If it is placed
too far into the program code a mode switch may occur before the task's
LO-mode budget is extended due to delays. We show in
Section~\ref{sec:checkpoint-location} the effects of using checkpoints at
different locations in a program's code.

\vspace{-4pt}
\subsection{\litmusrt{} Implementation}
\vspace{-4pt}
As a first step to applying \oursystem{} for use in adaptive
mixed-criticality scheduling, we extended \litmusrt{} with AMC
support. This required modifications to the existing partitioned
fixed-priority scheduling policy, to include the following new
variables in the task properties: \texttt{c\_{lo}},
\texttt{c\_{hi}}, \texttt{r\_{lo}}, \texttt{r\_{star}}, 
\texttt{c\_{extended}}, \texttt{max\_extended\_budget}.

Tasks are divided into low- and high-criticality classes. By default,
the HI-mode budget for low-criticality tasks, \texttt{c\_{hi}}, is set
to zero. If desired, we also support a reduced, non-zero HI-mode
execution budget for low-criticality tasks, as discussed in the work
on Imprecise Mixed-criticality
scheduling~\cite{liu2018scheduling}. Task priorities are determined
offline, along with all response times for a system operating in
LO-mode (\texttt{r\_{lo}}) and during a mode switch
(\texttt{r\_{star}}). These parameters are then initialized in the
kernel when the system starts executing a set of tasks.

The system is started in LO-mode, with all tasks assigned their
\texttt{c\_{lo}} budgets. Whenever a high-criticality task is out of
its LO-mode budget, an enforcement timer handler (based on Linux's
high-resolution timer~\cite{hrtimer}) is fired, and the system is
switched to HI-mode.

For AMC-\oursystem{}, \texttt{c\_{extended}} is the extended LO-mode
budget for a delayed job of a high-criticality task. An enforcement
timer handler is therefore triggered only when a high-criticality task
is still unfinished after the depletion of it \texttt{c\_{extended}}
time.

As Baruah et al. suggest~\cite{baruah2011response}, an AMC system
switches back to LO-mode when none of the high-criticality tasks have
been running for more than their LO-mode budgets. In the case of
AMC-\oursystem{}, the system will switch to a lower criticality level
when none of the high-criticality tasks have been running for more
than their extended LO-mode budgets. A list is used to keep track of
which high-criticality tasks complete within their (extended) LO-mode
budgets, to determine when to revert to a lower system criticality
level.

A task's LO-mode budget is extended by AMC-\oursystem{} by making
a \texttt{set\_rt\_task\_param} system call inside the
\texttt{ANNOUNCE\_TIME} macro. We have 
implemented the \texttt{task\_change\_params} callback of a 
\litmusrt{} \texttt{sched\_plugin} interface to support runtime adjustment of 
task parameters. In the \texttt{task\_change\_params} callback, the
system runs an online schedulability test. If the test returns
$true$, the budget extension is approved, and the enforcement timer
for the current task is adjusted accordingly. The
task's \texttt{c\_{extended}} variable is updated, along with the
maximum value of its extended budget
in \texttt{max\_extended\_budget}.

It is worth noting that a budget extension could be delayed until a
mode switch is about to happen. However, this strategy needs the
scheduler to save the delay at a checkpoint and later decide about the
budget extension at a mode switch point. This approach potentially
increases runtime overhead while repeatedly accumulating task delays.
Conversely, redundant budget extensions in \oursystem{} are avoided by
following a lazy budget extension approach. Such strategies are open
to further study.

%\section{Implementation Details}
%\input{implementation}

\vspace{-6pt}
\section{Evaluation}
\label{sec:evaluation}
\vspace{-6pt}
We tested our implementation of AMC and AMC-\oursystem{} in
\litmusrt{} with real-world
applications on an Intel NUC Kit~\cite{nuc}. The machine has an Intel
Core i7-5557U 3.1GHz processor with 8GB RAM, running Linux kernel 4.9.
We use three applications in our evaluations: a
high-criticality object classification application from the Darknet
neural network framework~\cite{darknet13}, a high-criticality object
tracking application from dlib C++ library \cite{dlib}, and a
low-criticality MPEG video decoder~\cite{ffmpeg18}.  These
applications are chosen for their relevance to the sorts of
applications that might be used in infotainment and autonomous driving
systems. The dlib object tracking application is only used for the
last set of experiments to test two different execution time
prediction models.

For object classification, we use the COCO
dataset~\cite{lin2014microsoftcoco} images for both profiling and
execution. The dataset for the Profiling phase was chosen from random images 
from the dataset and used to determine the LO- and HI-mode budgets of the 
Darknet high-criticality application. The dataset for the Execution phase was 
also randomly chosen but similarly distributed in terms of image dimensions as 
the data of the Profiling phase was.
For the video decoder application, we use the \emph{Big
Buck Bunny} video~\cite{bigbuckbunny} as the input. We have turned off
memory locking with \texttt{mlock} by the \texttt{liblitmus} library, as 
multiple object classification tasks collectively require more RAM than the 
physical 8GB 
limit.

\vspace{-4pt}
\subsection{Task Parameters}
\label{sec:eval_task_param}
\vspace{-4pt}
Table~\ref{tab:amc_taskset} shows the LO- and HI-mode budgets for the
three applications.  Each object classification and tracking task consist of a
series of jobs that classify objects in a single image. Each video
decoder task decodes 30 frames in a single job.

\vspace{-6pt}
\begin{table}[!ht]
	\footnotesize
	\centering
	\caption{Applications and their Budgets}
	\label{tab:amc_taskset}
	\vspace{-6pt}
	\renewcommand{\arraystretch}{1.15}
	\begin{subfigure}[t]{\textwidth}
		\begin{tabularx}{\textwidth}{|X|X|X|}
			\hline
			\textbf{Application} & \textbf{C(LO)} & \textbf{C(HI)} \\\hline\hline
			Darknet object classification & 345 ms & 627 ms \\\hline
			dlib object tracking & 10.7 ms & 18 ms \\\hline
			video decoder & 250 ms & - \\\hline
		\end{tabularx}
	\end{subfigure}\hfill
	\vspace{-4pt}
\end{table}
\vspace{-2pt}

The LO-mode budget, $C(LO)$, is the average time that a task takes to complete
its job. In Section~\ref{sec:overestimated_lo_budget}, we also present
experiments where we increase our LO-mode budget estimate. The HI-mode budget,
$C(HI)$, accounts for the worst-case running time of the high-criticality task
for any of its jobs. The LO-mode utilization of each individual task is
generated by the UUnifast algorithm~\cite{bini2005measuring}. We then
calculate a task's period by dividing its LO-mode budget by its utilization.

%Figure~\ref{fig:max_exec_cost} shows the maximum execution times of 10
%high-criticality object classification tasks in the presence of 10 other
%low-criticality video decoder tasks.
In all experiments, we observed that none of the tasks exceed their
HI-mode budget. Consequently, none of the high-criticality tasks miss their
deadlines in any of our tests. Hence, we assume that our derivation of LO- and 
HI-mode budgets are safe and correct for our experiments. The main focus of our 
evaluation now is to compare the QoS of the LC tasks.

\vspace{-4pt}
\subsection{QoS Improvements for Low-criticality Tasks}
\vspace{-4pt}
We compare the QoS for low-criticality tasks using AMC and 
AMC-\oursystem{} in different cases. In every case, each taskset has 
an equal number of high-criticality object classification tasks and 
low-criticality video decoder tasks. We experiment with ten 
schedulable tasksets in all cases except the base case described 
below. We run each of the tests ten times, and we report the average 
of the measurements for low-criticality tasks. As stated earlier, all 
high-criticality tasks meet their timing requirements in each case. 
Execution time prediction at a checkpoint for \oursystem{} is done 
with a linear extrapolation model for all the experiments 
($f(C_i(LO), X) = C_i(LO) + \ddfrac{C_i(LO) \times X}{100}$ from 
Section~\ref{sec:theory}) except in 
Section~\ref{sec:prediction_model} where we discuss other 
approaches.

Our base case is to run one high-criticality object classification task and
one low-criticality video decoder task for 180 seconds. Here, we set the
periods of both tasks to 1000 ms rather than using the UUnifast algorithm,
yielding a total LO-mode utilization of \textasciitilde 60\%.

\begin{figure}[!t]
%	\medskip
	\centering
	\RawFloats
	\footnotesize
	\caption{Video Decoder Performance}
	\label{fig:amc_lc_qos_2}
	\vspace{-10pt}
	\begin{subfigure}[t]{0.33\textwidth}
		\centering
		\includegraphics[width=0.97\textwidth]{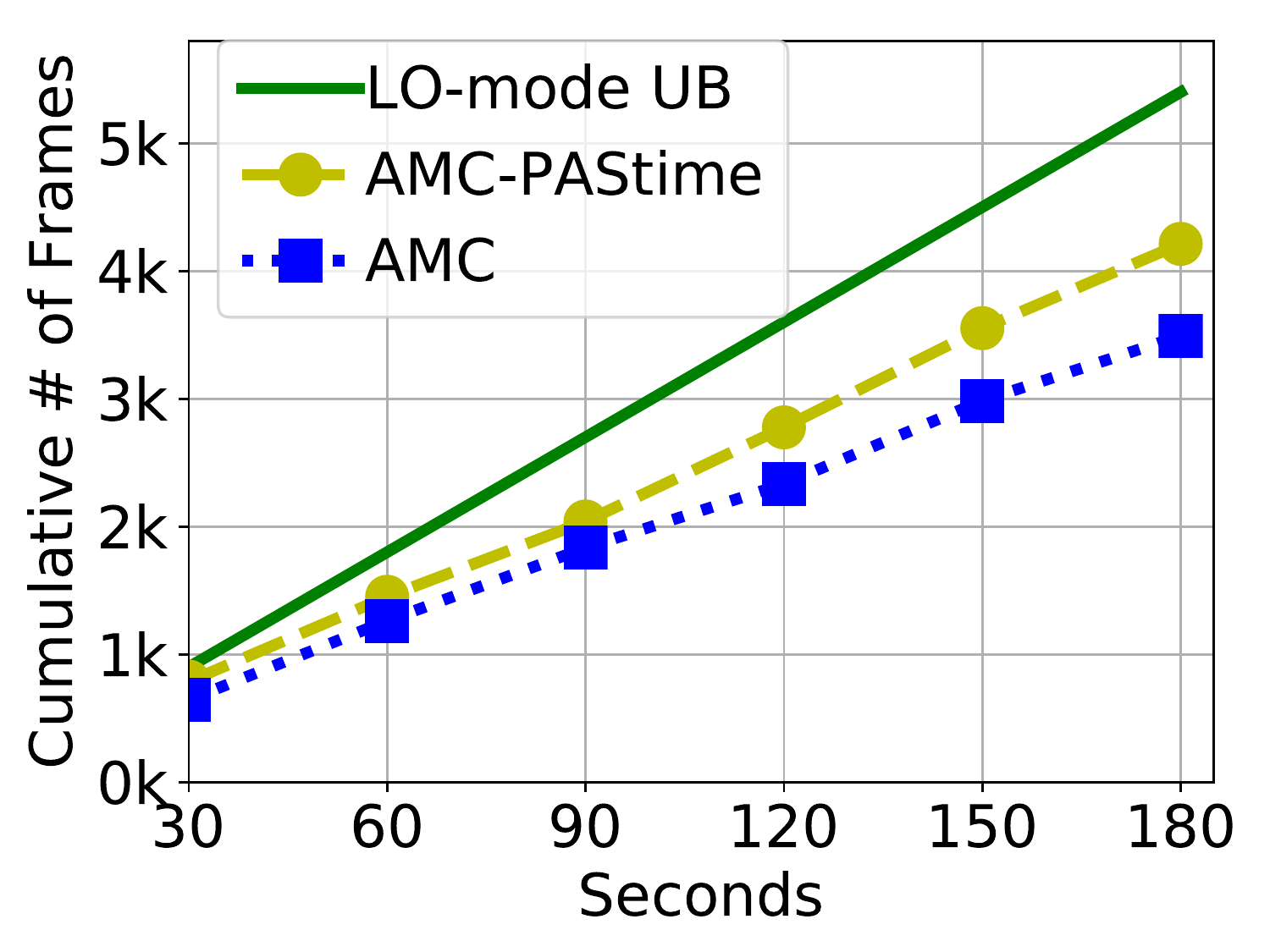}
		\vspace{-4pt}
		\hspace{2.0in}\caption{Cumulative \# of Frames}
		\label{fig:time_amc_cumul_frames}
%		\vspace{-16pt}
	\end{subfigure}\hfill
	\begin{subfigure}[t]{0.33\textwidth}
		\centering
		\includegraphics[width=0.99\textwidth]{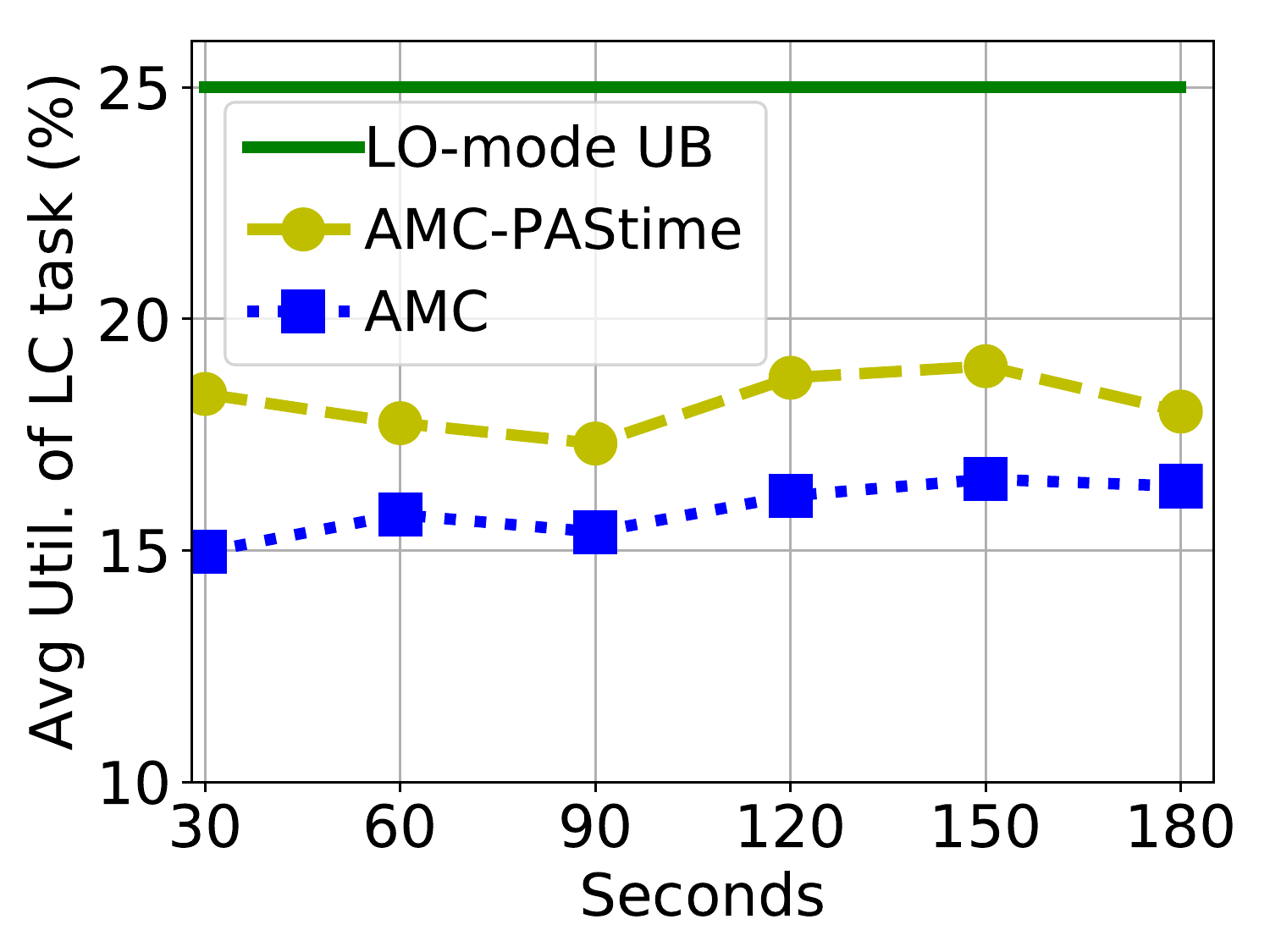}
		\vspace{-13pt}
		\hspace{0.25in}\caption{Average Utilization}
		\label{fig:avg_util_two_amc}
%		\vspace{-16pt}
	\end{subfigure}\hfil
	\begin{minipage}[t]{0.33\textwidth}
		\centering
		\vspace{-107.5pt}
		\caption{Varying \# of Tasks}
		\label{fig:task_amc_avg_util}
		\vspace{-10pt}
		\includegraphics[width=0.99\textwidth]{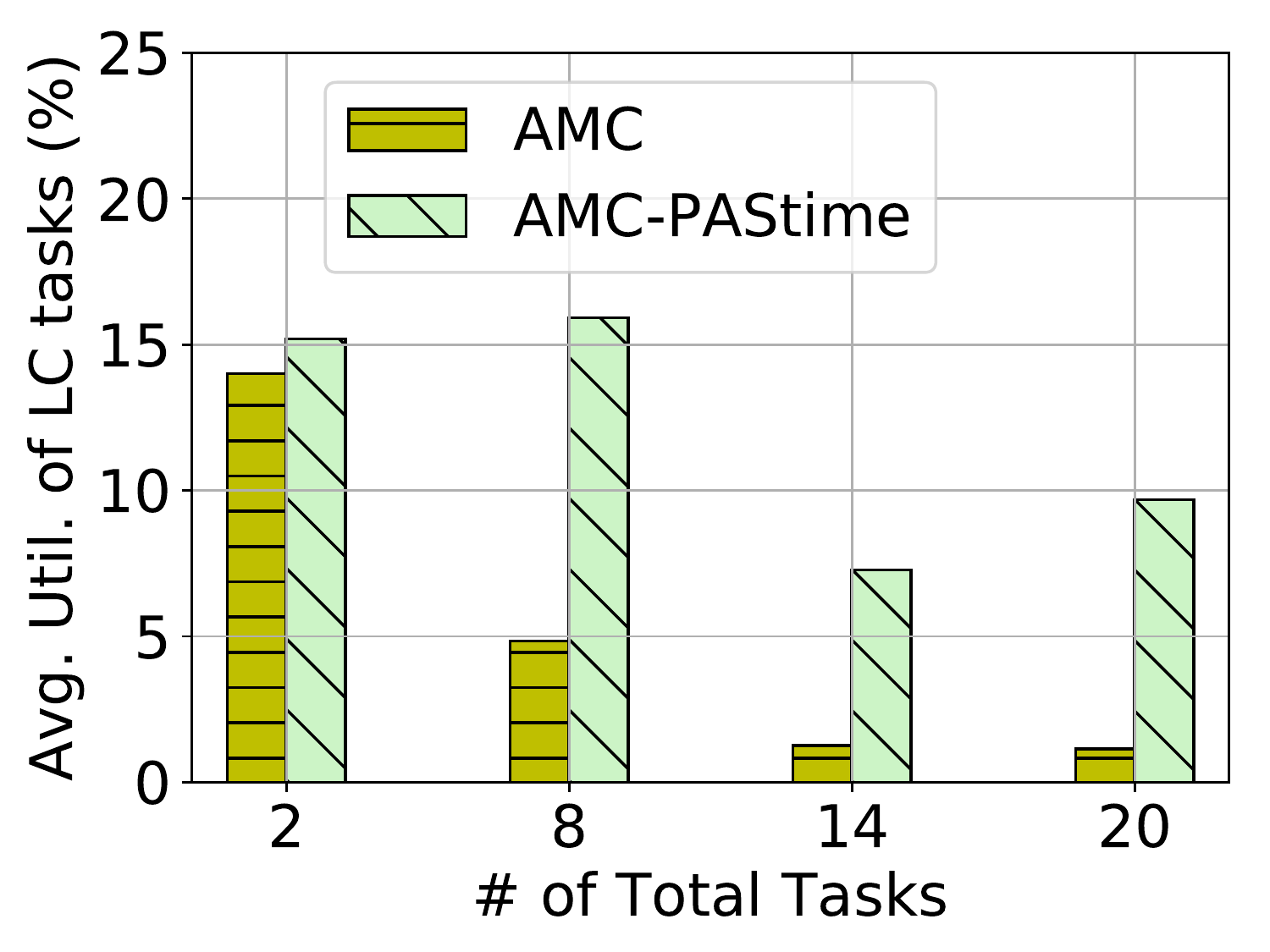}
	\end{minipage}
	\vspace{-4pt}
\end{figure}

Figure~\ref{fig:time_amc_cumul_frames} shows the cumulative number of frames
decoded by the video decoder task. The LO-mode Upper Bound (UB) line shows the
cumulative number of decoded frames if the system is kept in LO-mode all the 
time. This line represents a theoretical UB for a decoded 
frame playback rate of 30 frames per second over the entire experimental run.

We see in Figure~\ref{fig:time_amc_cumul_frames} that AMC-\oursystem{} has a
9--21\% increase in the cumulative number of decoded frames compared to AMC
scheduling.  The performance of the low-criticality task is related to the
number of HI-mode switches in the two scheduling policies.
%Figure~\ref{fig:time_amc_mode_switch} shows that AMC-\oursystem{} decreases
%the number of HI-mode switches by 35\% over the entire execution run, compared
%to AMC scheduling.
AMC-\oursystem{} decreases the number of HI-mode switches by 35\%, compared to 
AMC scheduling.

\footnotesize
\normalsize

Although the number of decoded frames is an illustrative metric for a video
decoder's QoS, the average utilization of an application is a more generic
metric. Average utilization represents the CPU share a task receives over a
period of time. Figure~\ref{fig:avg_util_two_amc} shows that AMC-\oursystem{}
achieves 10\% more utilization on average for the video decoder task, compared
to AMC scheduling. The LO-mode UB line is the maximum utilization of the video
decoder task, which is 25\% (i.e., $C(LO)/Period$ = 250 ms/1000 ms).

\vspace{-4pt}
\subsection{Scalability}
\vspace{-4pt}
To test system scalability, we increase the number of tasks in a taskset up to
20 tasks. As explained above, we generate the periods of the tasks by
distributing the total LO-mode utilization of \textasciitilde60\% to all the
tasks using the UUnifast algorithm.  This
setup is inspired by the theoretical parameters in previous mixed-criticality
research work~\cite{baruah2011response}. The LO-mode
utilization bound for low-criticality tasks remains between 25--35\%.

Figure~\ref{fig:task_amc_avg_util} shows the average utilization of the
low-criticality tasks, when the total tasks vary from 2 to 20. Each task in
this case consists of 20 jobs. We see that the average utilization drops for
AMC scheduling as the number of tasks increases.

AMC-\oursystem{} achieves significantly greater average utilization for the
low-criticality tasks, by deferring system switches to HI-mode until much
later than with AMC scheduling. This is because the LO-mode budgets for the
high-criticality tasks are extended due to runtime delays.

%Table~\ref{tab:amc_mode_switches_tasks} shows that 
AMC-\oursystem{} decreases
the number of mode switches by 28--55\%. AMC-\oursystem{}'s resistance to
switching into HI-mode allows low-criticality tasks to make progress. This in
turn improves their QoS. In these experiments, AMC-\oursystem{} improves the
utilization of the low-criticality tasks by a factor of $3$, $5$ and
$9$, respectively, for 8, 14 and 20 tasks. 
Table~\ref{tab:amc_mode_switches_tasks} shows that AMC-\oursystem{} 
reduces the number of mode-switches compared to AMC.

\begin{table}[t]
%	\captionsetup[subfigure]{justification=justified,singlelinecheck=false}
	\centering
	\footnotesize
	\caption{Number of Mode Switches}
	\label{tab:amc_mode_switches}
	\renewcommand{\arraystretch}{1.15}
	\vspace{-9pt}
	\begin{subfigure}[b]{0.44\columnwidth}
		\imagebox{2.9cm}{\begin{tabularx}{\textwidth}{|X|X|X|}
				\hline
				\textbf{\# of tasks} & \textbf{AMC} & 
				\textbf{AMC-\oursystem{}} \\\hline\hline
				2	& 4 & 2 \\\hline
				8	& 9 & 4 \\\hline
				14	& 7	& 5 \\\hline
				20	& 5	& 3 \\\hline
		\end{tabularx}}
		\vspace{-6pt}
		\caption{Varying Number of Tasks}
		\label{tab:amc_mode_switches_tasks}
	\end{subfigure}\hfill
	\begin{subfigure}[b]{0.54\columnwidth}
		\imagebox{2.9cm}{
			\begin{tabularx}{\textwidth}{|X|p{0.75cm}|X|}\hline
				\textbf{Utilization} (\%) & \textbf{AMC} & 
				\textbf{AMC-\oursystem{}} \\\hline\hline
				40	& 11 &	4 \\\hline
				50	& 11 &	4 \\\hline
				60	& 9	& 4	\\\hline
				70	& 10 &	5 \\\hline
				80	& 11 &	9 \\\hline
			\end{tabularx}}
		\vspace{-6pt}
		\caption{Varying Initial LO-mode Utilization}
		\label{tab:amc_8tasks_mode_switches}
	\end{subfigure}
	\vspace{-22pt}
\end{table}

\vspace{-4pt}
\subsection{Varying the Initial Total LO-mode Utilization}
\vspace{-4pt}
In this test, we vary the initial total LO-mode utilization for 8 tasks from
40\% to 80\% by adjusting the periods of all tasks. The initial utilization
does not account for increases caused by LO-mode budget extensions to
high-criticality tasks.
%
%\begin{figure}[t]
%%	\medskip
%	\centering
%	\RawFloats %For Minipage
%	
%	\begin{minipage}[t]{0.49\columnwidth}
%		\centering
%		\caption{Varying LO-utilization}
%		\label{fig:util_amc_avg_lc_util}
%		\vspace{-10pt}
%		\includegraphics[width=0.85\textwidth]{./data/util_amc_avg_lc_util}
%	\end{minipage}
%	\vspace{-18pt}
%\end{figure}

Figure~\ref{fig:util_amc_avg_lc_util} demonstrates
that AMC-\oursystem{} improves average utilization of the low-criticality
tasks by more than 3 times, up to 70\% total LO-mode utilization. After that,
AMC-\oursystem{} and AMC scheduling converge to the same average utilization
for low-criticality tasks. This is because there is insufficient surplus CPU
time in LO-mode for AMC-\oursystem{} to accommodate the extended budget of a
high-criticality task. Therefore, the LO-mode extension requests are
disapproved by AMC-\oursystem{}. The reduced number of mode switches for 
AMC-\oursystem{}, shown in Table~\ref{tab:amc_8tasks_mode_switches}, also 
corroborates the rationale behind AMC-\oursystem{}'s 
better performance than AMC.

%Table~\ref{tab:amc_8tasks_mode_switches} demonstrates that AMC-\oursystem{}
%improves the low-criticality tasks' QoS by reducing the number of mode
%switches by \textasciitilde50\% until 70\% LO-mode utilization.

We note that the schedulability of random tasksets decreases with higher
LO-mode utilization in AMC scheduling. Therefore, many real-world tasksets may
not be schedulable because of their HI-mode utilization. Thus,
AMC-\oursystem{}'s improved performance is significant for practical
use-cases.
%\vspace{-2pt}

\vspace{-4pt}
\subsection{Estimation of LO-mode Budget}
\label{sec:overestimated_lo_budget}
\vspace{-4pt}
In our evaluations until now, we estimate a LO-mode budget based on the
average execution time of the high-criticality object classification
application. In the next set of experiments, we estimate the LO-mode budget of
a high-criticality task to be a certain percentage above the average profiled
execution time. An increased LO-mode budget for high-criticality tasks
benefits AMC scheduling. This is because high-criticality tasks are now given
more time to complete in LO-mode, and therefore low-criticality tasks will
still be able to execute as well. As a result, the utilization of
low-criticality tasks is able to increase.

Suppose that $C(LO)$ is an average execution time estimate for the
LO-mode budget of a high-criticality task. Let $(C(LO) + o)$ be an
overestimate of the LO-mode budget. As before, AMC-\oursystem{}
detects an $X\%$ lag at a checkpoint and predicts the total execution
time to be $C(LO) + e$. If the actual LO-mode budget is $(C(LO) +
o)$ then AMC-\oursystem{} requests for an extra budget of $(e-o)$,
assuming $(e-o) > 0$. 
%As will be clear in
%Section~\ref{sec:prediction_model}, AMC-\oursystem{} supports other
%prediction models to calculate $e$.

\begin{figure}[!t]	
	\centering
	\RawFloats
	%	\medskip
	\begin{minipage}[t]{0.33\columnwidth}
		\centering
		\caption{Varying LO-util.}
		\label{fig:util_amc_avg_lc_util}
		\vspace{-10pt}
		\includegraphics[width=0.98\textwidth]{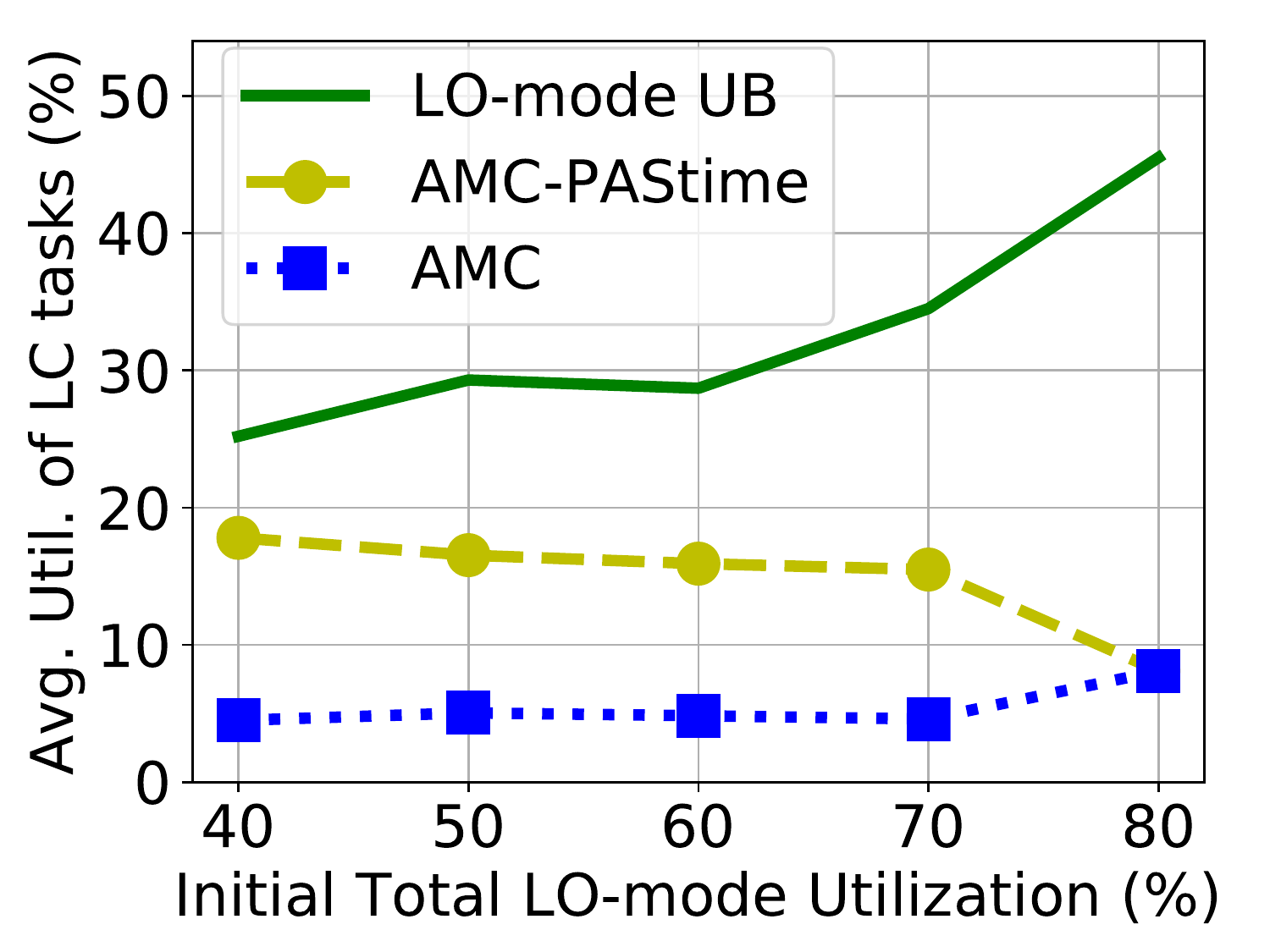}
	\end{minipage}\hfil
	\begin{minipage}[t]{0.33\columnwidth}
		\centering
		\caption{Overestimated $C(LO)$}
		\label{fig:amc_overestimate}
		\vspace{-10pt}
		\includegraphics[width=0.98\textwidth]{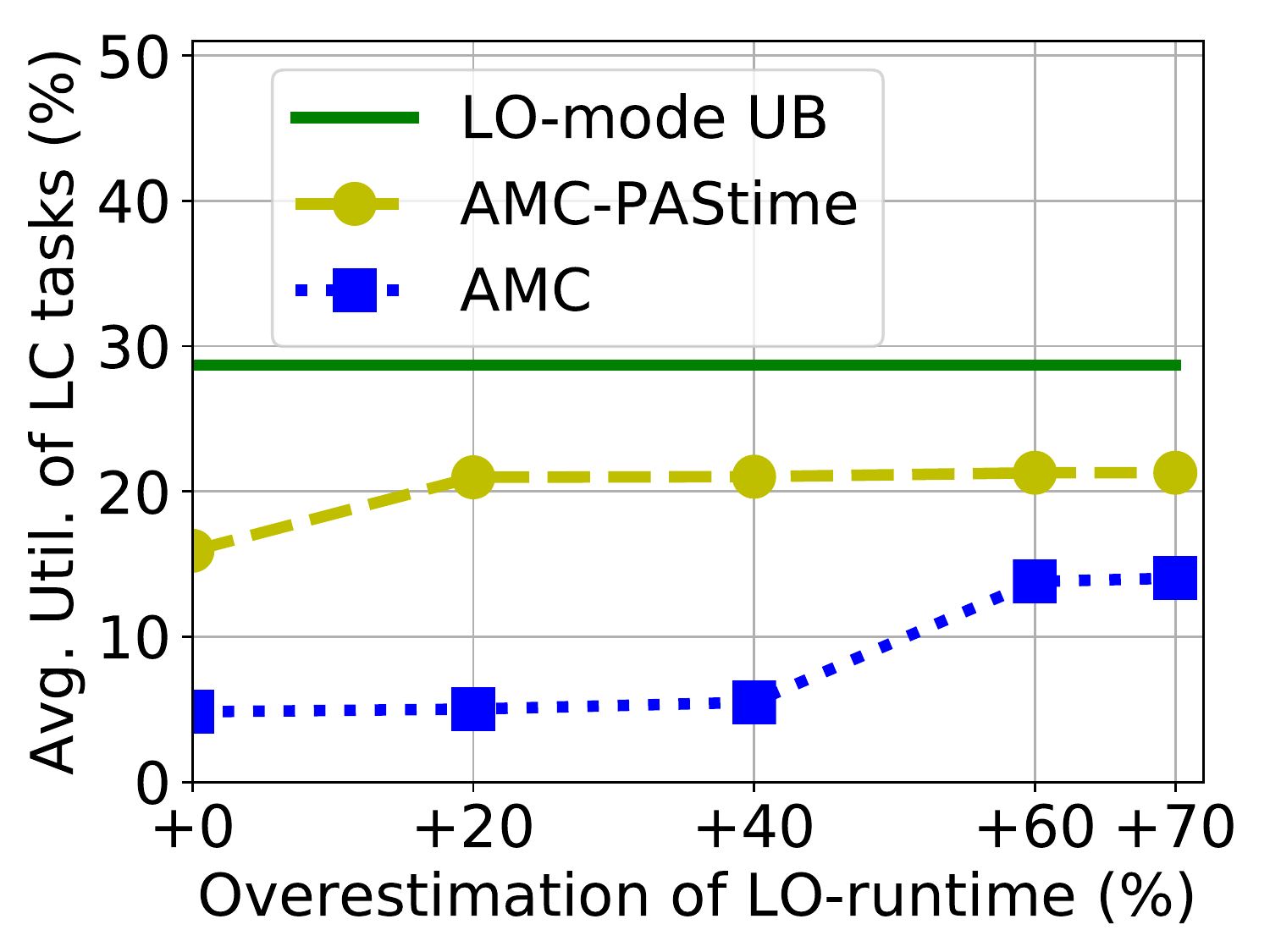}
	\end{minipage}\hfil
	\begin{minipage}[t]{0.33\textwidth}
		\centering
		\caption{Checkpoint Location}
		\label{fig:checkpoint_location_amc}
		\vspace{-10pt}
		\includegraphics[width=\textwidth]{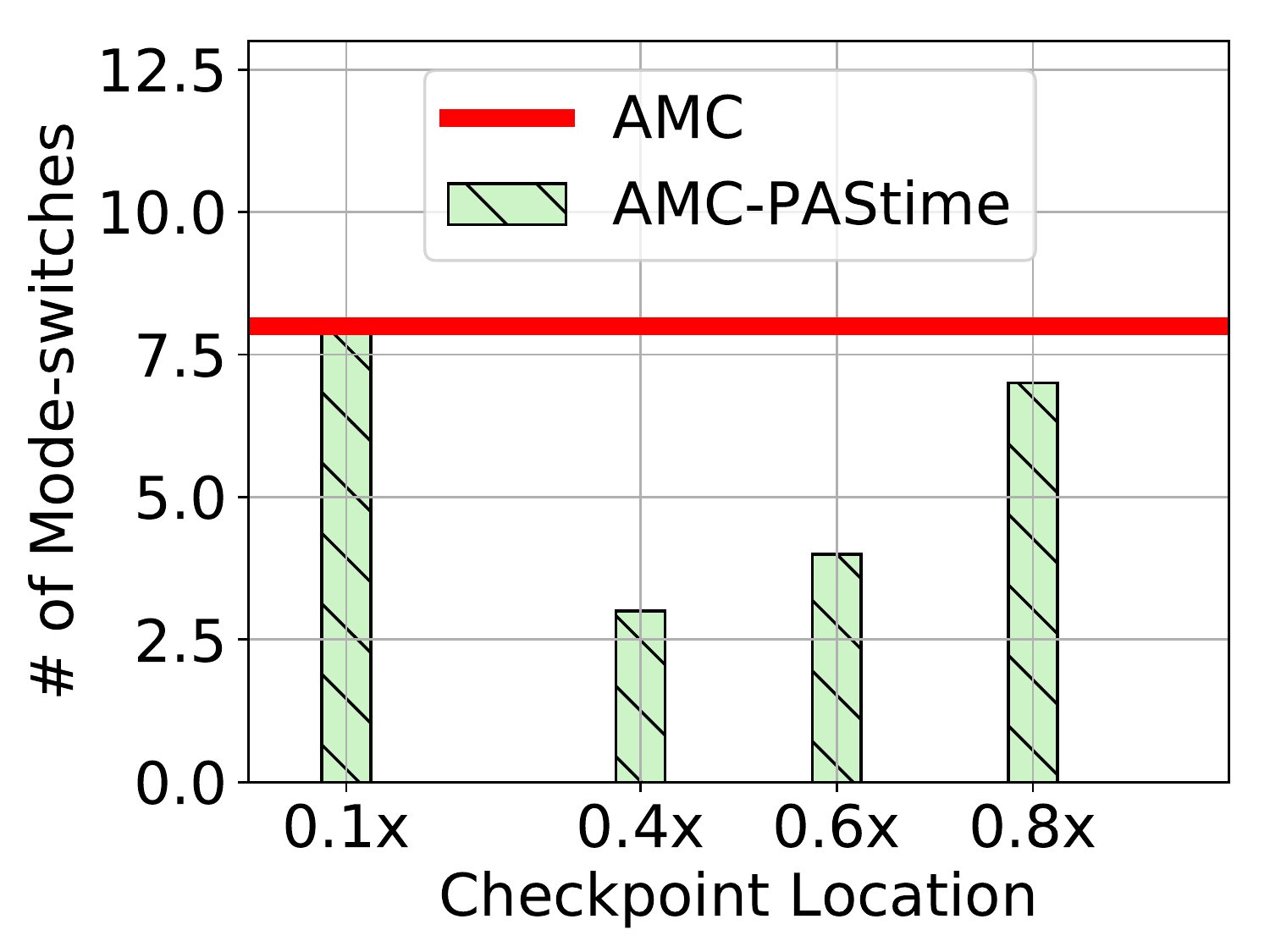}
	\end{minipage}
	\vspace{-4pt}
\end{figure}

Even for overestimated $C(LO)$, Figure~\ref{fig:amc_overestimate} shows that 
AMC-\oursystem{} still improves
utilization for the low-criticality tasks by more than a factor of 3 up to an
overestimation of 40\%. Overestimation helps in reducing the number of mode
switches for AMC scheduling after 40\%, as high-criticality tasks have larger
budgets in LO-mode.

There is no improvement by AMC scheduling after 60\% LO-mode budget
overestimation.  AMC-\oursystem{} also shows no benefits with increased
overestimation, because the LO-mode budget extensions are disapproved by the
online schedulability test. Therefore, the system is switched to HI-mode by an
overrun of a high-criticality task.

\vspace{-4pt}
\subsection{Checkpoint Location}
\label{sec:checkpoint-location}
\vspace{-4pt}
We use our modified LLVM compiler in the Profiling phase to determine a viable 
checkpoint for the high-criticality object classification task. We instrument 
checkpoints in the 
\texttt{forward\_network} function of the Darknet neural network module. 
We consider four checkpoint locations in the Profiling phase, which are both
automatically and manually instrumented. In the Execution phase, we measure
performance for each of these checkpoint
locations. Figure~\ref{fig:checkpoint_location_amc} shows the variation in the
number of mode switches against the location of a checkpoint. The x-axis is
the approximated division point of a checkpoint location with respect to
$C(LO)$. For example, 0.1$\times$ means that the checkpoint is at $(0.1
\times 
C(LO))$.

We see that the number of mode switches decreases if the location of a
checkpoint is more towards the middle of the code. However, a checkpoint near
the start and the end of the source code have nearly the same number of mode
switches, as with AMC scheduling. A checkpoint near the beginning of a program
is not able to capture sufficient delay to increase the LO-mode budget enough
to prevent a mode switch. Likewise, a checkpoint near the end of a program is
often too late. A HI-mode switch may occur before the high-criticality task
even reaches its checkpoint. Hence, a checkpoint at $0.8\times$ in the source
code of a program reduces the number of mode switches by just 1.

\vspace{-4pt}
\subsection{Overheads}
\vspace{-4pt}
The main overheads of AMC-\oursystem{} compared to AMC are the online
schedulability test and budget extension. We first derive an upper bound on the 
overhead by offline analysis and compare with the experimental measurements.

Our offline upper bound is the total number of iterations in solving the
response time recurrence relations during the schedulability test in
AMC-\oursystem{}.  We generate 500 random tasksets of 20 tasks for different
initial LO-mode utilizations. Initial LO-mode utilizations range from 40\% to
90\%. The utilization of each individual task is generated using the UUnifast
algorithm, and each period is taken from 10 to 1000 simulated time units, as
done in previous
work~\cite{baruah2011response,huang2014implementation}. As our experimental
taskset has a criticality factor (CF~$=\ddfrac{C(HI)}{C(LO)}$)
of \textasciitilde1.8, we also test with a CF of $1.8$.

Among the schedulable tasks with AMC scheduling, we increase the demand in
LO-mode budget of the highest priority task.  Then, we calculate the total
number of iterations needed to determine whether an extension of the budget
can be approved by an offline version of
the online schedulability test.
%Algorithm~\ref{algo:change_budget_runtime}. 
Here, one iteration is a single
update to the response time in any one of the recurrence equations (in
Equations~\ref{eq:LO_AMC_Online} and ~\ref{eq:HC_AMC_Star_Online}) used to
test for schedulability.

We increase the demand for extra budget from 10 to 80\% in this
analysis because the CF is $1.8$. We check the maximum number of
iterations to decide the schedulability of a taskset across the 500
tasksets. We carry out this offline analysis with 40-90\% initial
LO-mode utilization.

\begin{figure}[!t]	
	\centering
	\RawFloats
	\begin{minipage}[t]{0.33\columnwidth}
		\centering
		\caption{Offline Iterations}
		\label{fig:offline_iterations}
		\vspace{-8pt}
		\includegraphics[width=0.95\textwidth]{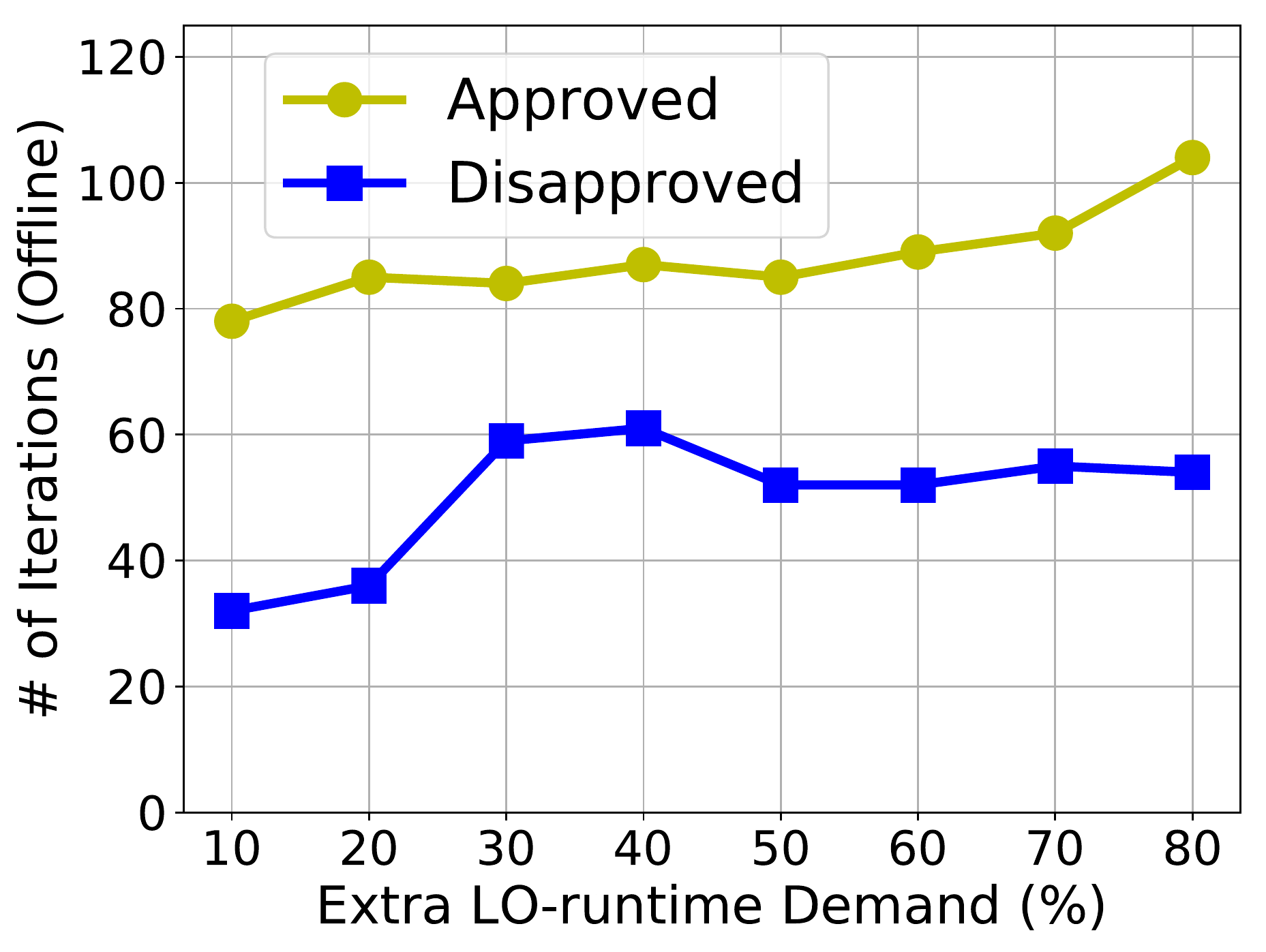}
	\end{minipage}\hfil
	\begin{minipage}[t]{0.34\textwidth}
		\centering
		\caption{Online Iterations}
		\label{fig:iterations}
		\vspace{-10pt}
		\includegraphics[width=0.98\textwidth]{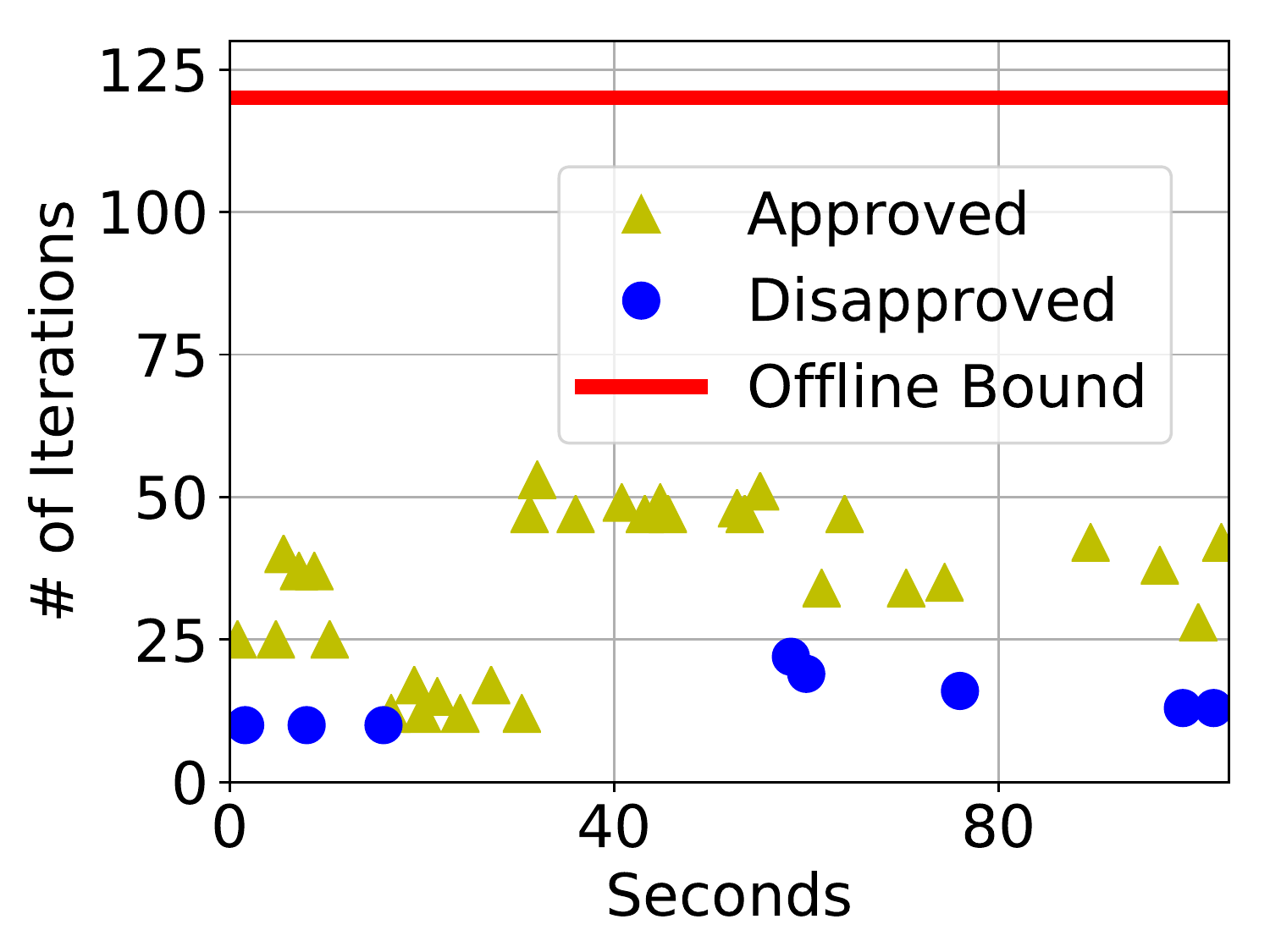}
	\end{minipage}\hfil
	\begin{minipage}[t]{0.33\textwidth}
		\centering
		\caption{Budget Extension}
		\label{fig:overheads}
		\vspace{-10.5pt}
		\includegraphics[width=\textwidth]{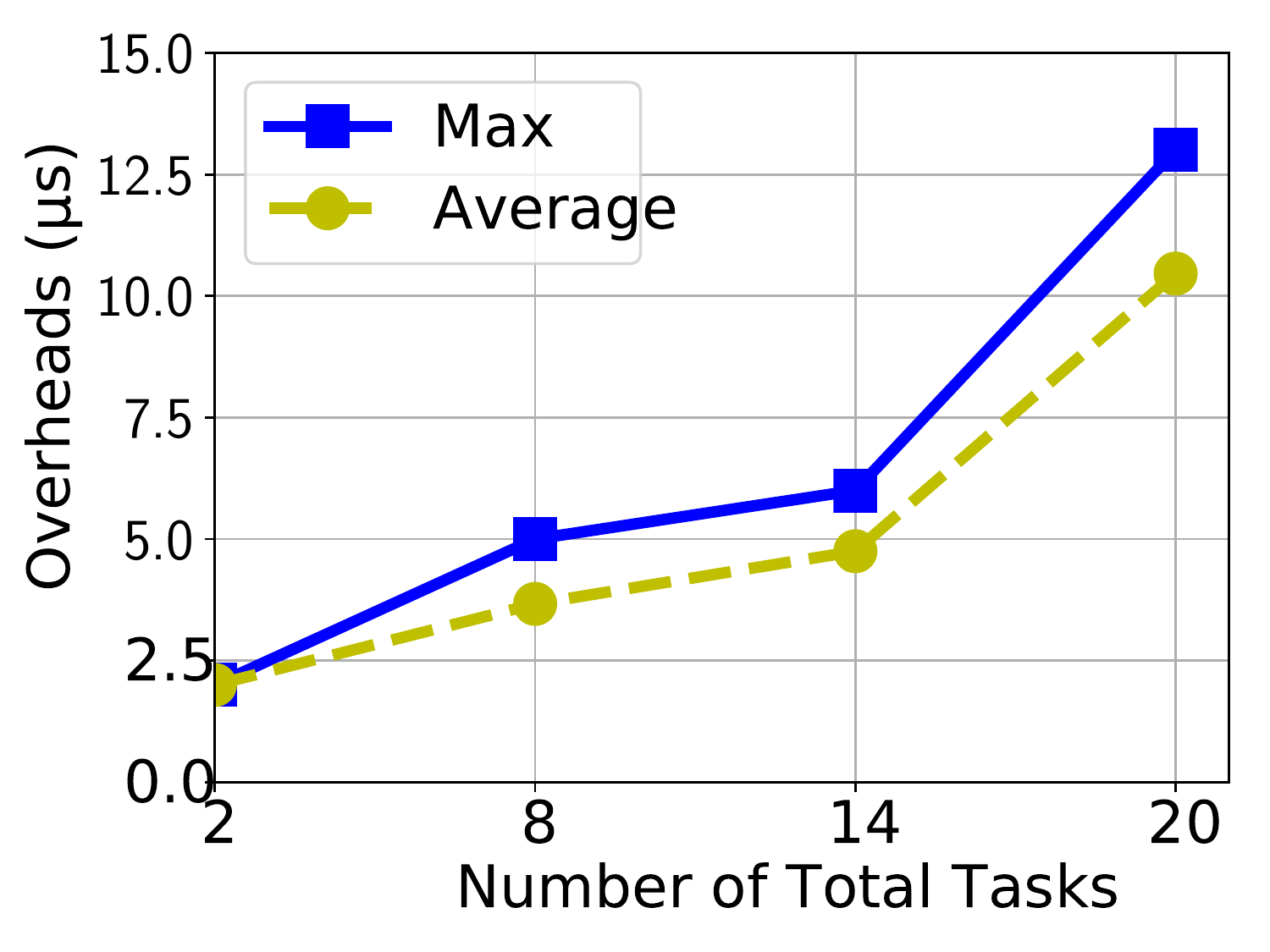}
	\end{minipage}\hfil
	\vspace{-14pt}
\end{figure}

In Figure~\ref{fig:offline_iterations}, we show the maximum number of
iterations to decide the schedulability of a taskset, against a demand of
10 to 80\% extra budget in LO-mode by the highest priority task. Each point in
the figure represents the maximum iterations across the 500 tasksets to either
approve or disapprove of schedulability.

%The figure shows the case with 60\% initial total LO-mode utilization (before
%applying budget extensions).  The extra demand shown on the x-axis cannot go
%beyond 80\% as the CF is $1.8$. As expected, disapproval takes fewer
%iterations than approval of a taskset.

We have observed the number of iterations to be
as high as $120$. Therefore, we set 120 as the highest number of allowed
iterations for the online schedulability test. When the number of iterations
exceed 120 at runtime, we disapprove a LO-mode budget extension. This strategy
maintains a safe and known upper bound on the online overhead of
AMC-\oursystem{}.

\vspace{-4pt}
\subsubsection{Microbenchmarks}
\vspace{-4pt}
Each iteration of a response time calculation has a worst-case time of
1$\mu$s, for our implementation of the
online schedulability test
%Algorithm~\ref{algo:change_budget_runtime}
in \litmusrt{}. Therefore, we bound the worst-case delay for the online
schedulability test in \litmusrt{} at 120$\mu$s for our test cases.
Additionally, the worst-case execution time for the
\texttt{ANNOUNCE\_TIME} macro is 10$\mu$s. Hence, the maximum total 
overhead of an \texttt{ANNOUNCE\_TIME} call, accounting for the
schedulability test, is 130$\mu$s. The 130$\mu$s overhead is factored into the
LO-mode extension inside the \litmusrt{} kernel.

Figure~\ref{fig:iterations} shows the number of iterations for the online
schedulability test for a taskset of 20 tasks with 60\% initial total LO-mode
utilization, in cases where an extension was approved and disapproved. We see 
that the offline bound of 120 iterations is much higher
than the actually observed number of iterations. Hence, we never need to
abandon the schedulability test because of excessive overheads. In addition,
disapproval takes less time than approval online, which corroborates our
offline observations in Figure~\ref{fig:offline_iterations}.

Figure~\ref{fig:overheads} shows the maximum and average times for a LO-mode
budget extension decision including the online schedulability test. It 
demonstrates that the extension approval decision takes 
more time with increasing number of tasks. However, the maximum times are still 
significantly lower than the offline upper bound of 130$\mu$s for 20 tasks. In 
general, budget extension overheads can be bounded according to the number of 
tasks in the system.

\vspace{-4pt}
\subsection{Execution Time Prediction Model}
\label{sec:prediction_model}
\vspace{-4pt}
As we have explained in Section~\ref{sec:theory}, the execution time
after a checkpoint is predicted by a function $f(C_i(LO), X)$, where
$C_i(LO)$ is the LO-mode budget, and $X\%$ is the observed delay
percentage relative to the LO-mode time to reach the checkpoint.  The parameter 
$X$ in $f$ is a {\em timing progress
metric}, which is used to make runtime scheduling decisions in
\oursystem{}.

%\footnotesize
%\begin{figure}[t]
%	\centering
%	\medskip
%	\RawFloats
%	
%	\begin{minipage}[t]{0.49\textwidth}
%		\caption{Prediction accuracy}
%		\label{fig:amc_pastime_prediction}
%		\vspace{-10pt}
%		\includegraphics[width=0.85\textwidth]{./data/prediction}
%	\end{minipage}
%	\vspace{-22pt}
%\end{figure}
%\normalsize

\vspace{-4pt}
\subsubsection{Linear Extrapolated Delay}
\vspace{-4pt}
We have already shown in the previous experiments how a
straightforward linear extrapolated delay improves the utilization
of LC tasks compared to AMC for an object classification
application. In such a case, $f(C_i(LO), X) = C_i(LO)
+ \frac{C_i(LO) \times X}{100}$. A linear extrapolation of delay at a
checkpoint applied to the entire LO-mode time makes sense in
the absence of additional knowledge that could influence the
remaining execution time of the task.
%% The linear extrapolation based on the
%% $X\%$ delay is effective because $X$ is calculated compared to the
%% LO-mode time to reach a checkpoint, $C_i^{CP}(LO)$.
%% Hence, $X$
%% effectively reflects the lag of the current job from the LO-mode
%% progress of the task until the checkpoint. In case a checkpoint's lag
%% from the LO-mode progress has a linear relationship with the lag from
%% the full task's LO-mode progress, the linear extrapolation is an
%% effective strategy. In other words, a checkpoint should be
%% representative of a full task's execution progress for linear
%% extrapolation to work.

We now investigate further whether the linear extrapolation is effective in 
detecting the amount of delay in our Darknet high-criticality object 
classification task. We compare the predicted and actual times taken by the 
high-criticality tasks when LO-mode is extended by
AMC-\oursystem{}. This experiment is performed with 8 tasks for 40 jobs each.
The initial total LO-mode utilization is 60\%, before applying budget
extensions.

\begin{figure}[t]
	\centering
	\RawFloats %For Minipage
	\begin{minipage}[t]{0.33\textwidth}
		\caption{Prediction Accuracy}
		\label{fig:amc_pastime_prediction}
		\vspace{-10pt}
		\includegraphics[width=0.99\textwidth]{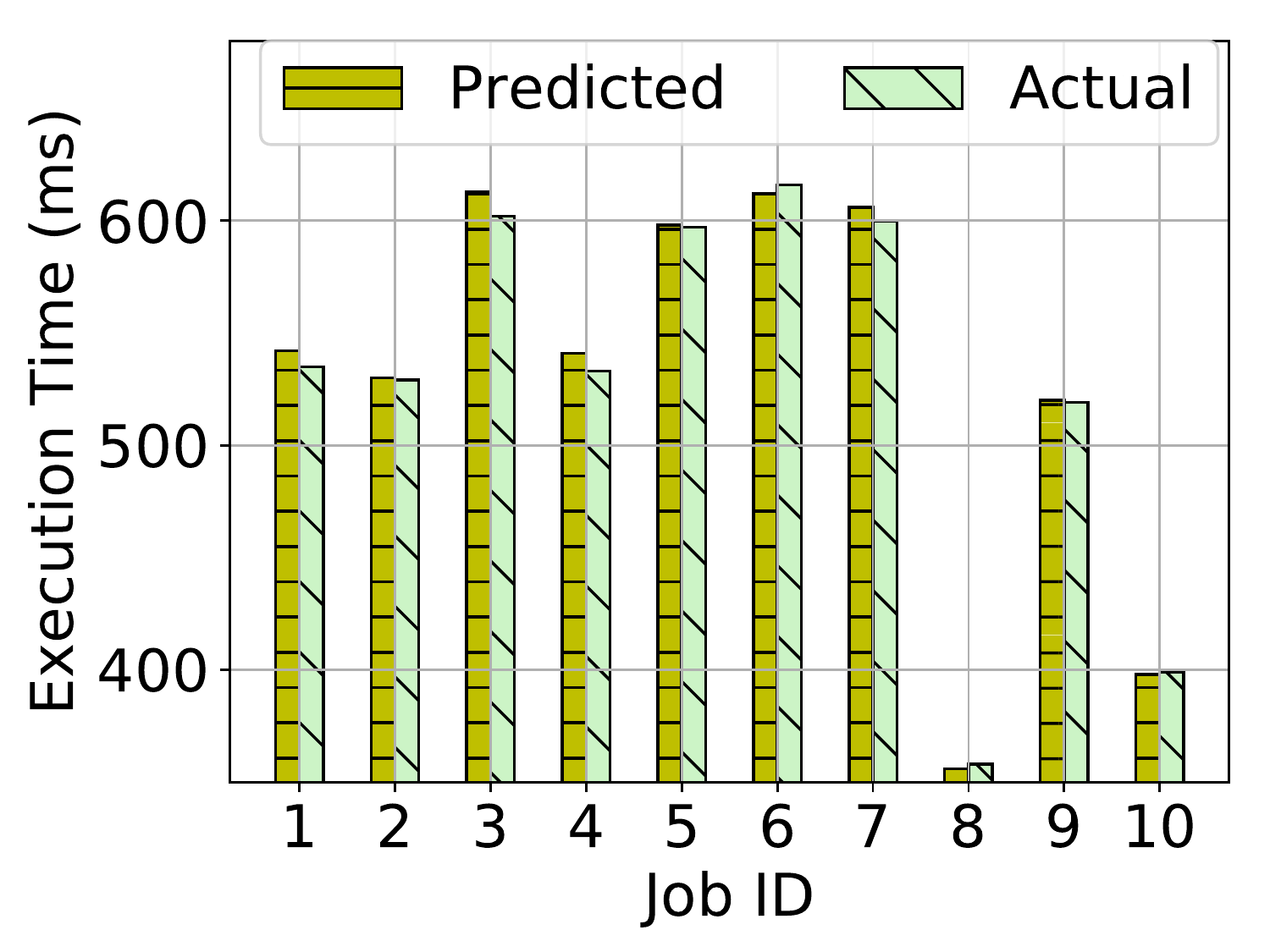}
	\end{minipage}\hfil
	\begin{minipage}[t]{0.33\textwidth}
		\centering
		\footnotesize
		\caption{Linear Extrapolation}
		\label{fig:linear_extrapolation}
		\vspace{-9pt}
		\includegraphics[width=0.94\textwidth]{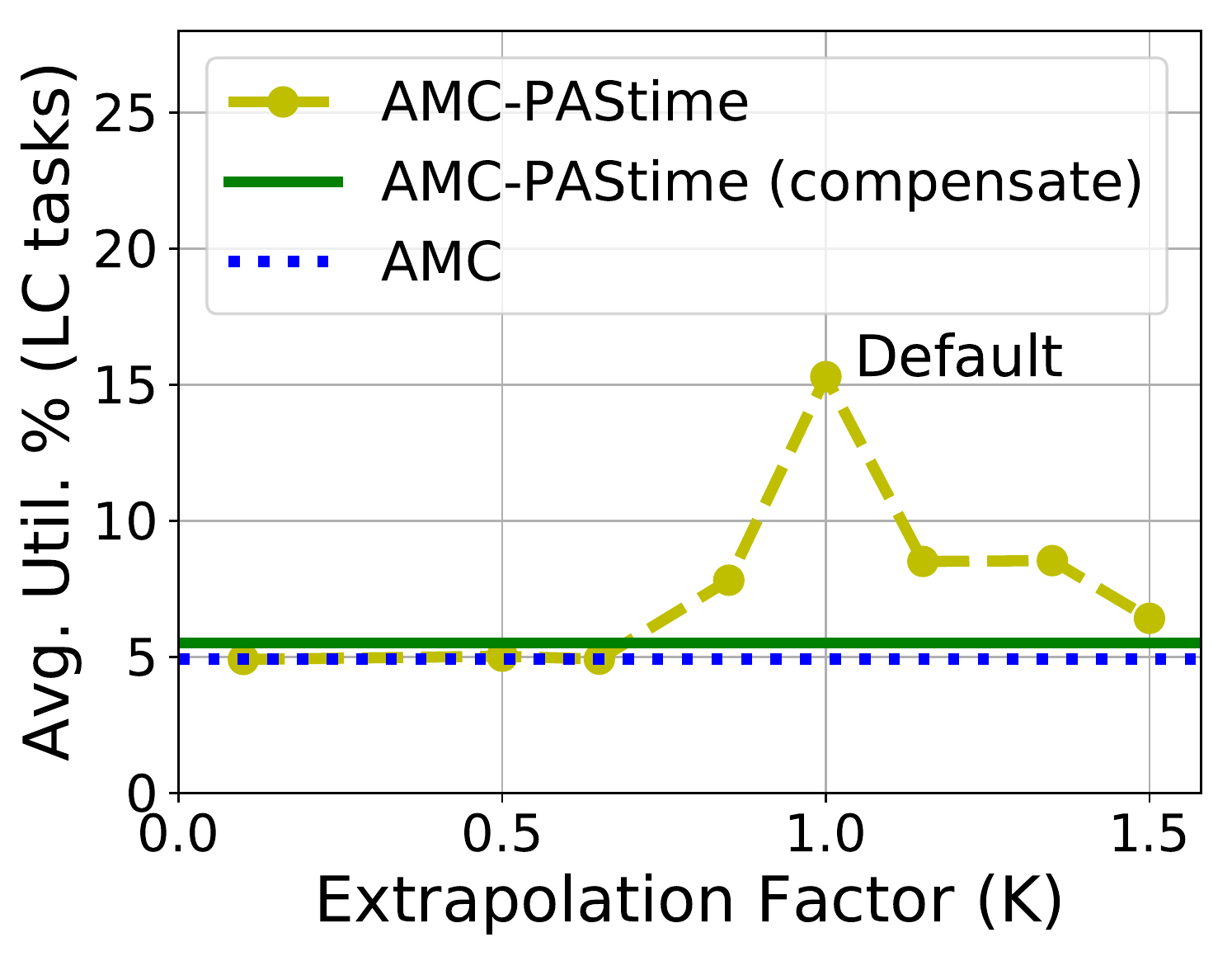}
	\end{minipage}\hfill
	\begin{minipage}[t]{0.32\textwidth}
		\centering
		\caption{Prediction Models}
		\label{fig:mem_performance}
		\vspace{-10pt}
		\includegraphics[width=\textwidth]{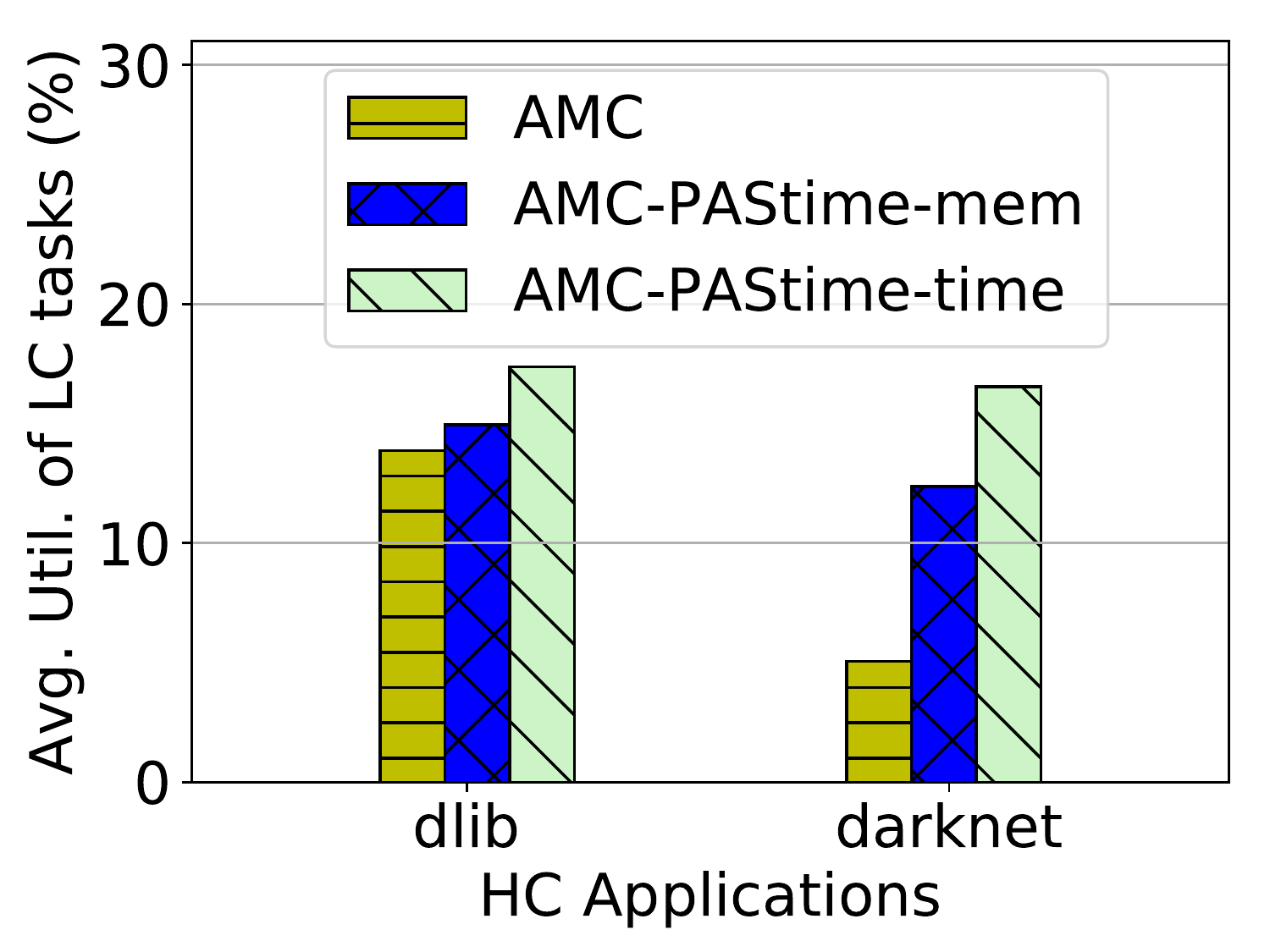}
	\end{minipage}
	\vspace{-4pt}
\end{figure}

We show ten of the extended jobs in
Figure~\ref{fig:amc_pastime_prediction}. We see that the predicted
execution times are close to the actual execution times. For cases
where the predicted times exceed the actual budget expenditure, the
predictions overestimate execution by just 0.88\% on average for this
experiment. In Figure~\ref{fig:amc_pastime_prediction}, Job ID 6, 8
and 10 show lower predicted times than the actual spent budgets. For
these jobs, the system is switched to HI-mode because the extended
LO-mode is not enough for a task to complete its job. In these cases,
the predicted times are smaller than the spent budgets by 0.49\%.

This experiment shows that the checkpoint is effectively being used to
predict the execution time of a high-criticality task in most
cases. The budget extensions are a reasonable estimate of the actual
task requirements.

\vspace{-4pt}
\subsubsection{Alternative Execution Time Compensation Models}
\vspace{-4pt}
In our experiments, we compare the linear extrapolation model with an
alternative compensatory execution time prediction model. The
alternative approach compensates for the observed delay at a
checkpoint, by adding the delay amount to the predicted execution
time. If the observed delay is $Y$, then the estimated total execution
time is $C^\prime_i(LO) = C_i(LO) + Y$.

We have also investigated variations to the linear extrapolation
model, even though it has proven effective in our previous
experiments. A further experiment multiplies the extrapolated delay by
a factor $K$, such that the predicted execution time is
$f(C_i(LO), X) = C_i(LO) + K \times \ddfrac{C_i(LO) \times X}{100}$.

Figure~\ref{fig:linear_extrapolation} compares AMC-\oursystem{} with
the alternative compensatory model {\tt compensate}, and linear
extrapolation model where $K$ ranges from 0.1 to 1.5. The linear
extrapolation model ($K=1$) outperforms the other approaches, while
the compensatory model improves the utilization slightly compared to AMC.

\vspace{-4pt}
\subsubsection{Prediction based on Memory Access Time}
\vspace{-4pt} Memory accesses by different processes compete for
shared cache lines and consequently cause unexpected
microarchitectural delays. There are previous works that model the
memory \cite{pellizzoni2011predictable,
  mancuso2014light,yun2014palloc} and cache
\cite{west2010online,ye2014coloris} accesses in a multicore machine to
deal with the issue of predictable execution. Here, we demonstrate
\oursystem{} with an execution time prediction model based on the
number of memory accesses, to increase the LO-mode budget of a
high-criticality task.

In this model, we note the number of memory accesses by a
high-criticality task as we measure the LO- and HI-mode execution
time of the task during the Profiling phase. We measure the average
number of memory accesses in LO-mode for each profiled run of a task
with different inputs, and denote it by $M(LO)$.

During the Profiling phase, we also measure the number of memory
accesses in LO-mode before and after a checkpoint:
$M_{\text{pre\_}CP}(LO), M_{\text{post\_}CP}(LO)$. These values are
also averaged across all runs of a given task. Therefore,
$M_{\text{pre\_}CP}(LO) + M_{\text{post\_}CP}(LO) = M(LO)$

We define a new {\em memory instructions progress metric} to be the ratio
of task execution time to the number of memory accesses. The LO-mode
value of this metric is $Pr_{mem, LO} =
\frac{C(LO)}{M(LO)}$.

In the Execution phase, we measure the number of memory accesses and
the used budget up to a checkpoint, respectively denoted by $M_{CP}$
and $C_{used}$ for a given task. During Execution, we calculate the memory 
instructions progress metric at a checkpoint, $Pr_{mem, CP}
= \frac{C_{used}}{M_{CP}}$.
If $Pr_{mem, CP} \le Pr_{mem, LO}$, we predict the task is progressing
as expected in LO-mode in terms of memory access-related delays. There
are other factors such as I/O-related delays that affect the execution
time of a task. However, I/O should be budgeted separately from the
main task, as shown in prior work for
AMC~\cite{missimer2016mixed}. Notwithstanding, \oursystem{} is capable
of supporting even richer prediction models based on a multitude of
factors that cause delays due to microarchitectural and task
interference overheads.

When $Pr_{mem, CP} > Pr_{mem, LO}$, we predict a task's
memory access delays. We calculate the expected number of
memory instructions after a checkpoint: $M_{\text{expected\_post\_CP}}
=
\frac{M_{CP} \times 
M_{\text{post\_CP}}(LO)}{M_{\text{pre\_}CP}(LO)}$, and 
increase the LO-mode budget by $(M_{\text{expected\_post\_}CP} 
\times (Pr_{mem, CP}$ - $Pr_{mem, LO}))$. This increment helps 
cover future memory delays after a checkpoint.

We tested both high-criticality dlib object tracking and Darknet
object classification applications with this
model. Figure~\ref{fig:mem_performance} shows experimental results
with 4 HC tasks (either dlib or Darknet) and 4 LC video decoder tasks,
with initial 50\% LO-mode utilization. We see that {\tt
  AMC-\oursystem{}-mem} with this memory access progress metric
provides better utilization to the LC tasks than AMC. However, it is
not better than {\tt AMC-\oursystem{}-time}, which uses the default
linear extrapolated execution time prediction. Nevertheless, the
result shows that some form of progress metric dynamically improves
the utilization of LC tasks.

%When we observe lag both in memory access and timing, then we decide 
%the 
%LO-mode budget based on only timing. The timing lag already 
%encapsulates the 
%delays related to memory instructions until a checkpoint.

%In case $(Pr_{mem, CP} - Pr_{mem, LO})$ is lesser than $\delta$ and there is 
%some positive delay $X\%$ at a checkpoint, then the task is lagging compared 
%to 
%LO-mode progress because of non-memory-related factors such as inputs to a 
%task. In such a case, we increase the LO-mode budget of the task by following 
%our linear extrapolation model. In case, $(Pr_{mem, CP} - Pr_{mem, LO})$ is 
%greater than $\delta$, then we increase the LO-mode budget by a 
%factor of $max((Pr_{mem, CP} - Pr_{mem, LO}), \frac{X}{100})$. In all these 
%scenarios, $\delta$ should be empirically decided.

\vspace{-4pt}
\section{Related Work}
\label{sec:related}
\vspace{-4pt}
The problem of determining tight worst-case execution time (WCET)
bounds for tasks~\cite{wilhelm2008worst} is compounded by timing
variations caused by caches, buses and other hardware
features. Recently, mixed-criticality systems
(MCSs)~\cite{vestal2007preemptive,baruah2008schedulability,baruah2010towards,de2009scheduling,baruah2011response,
burns2013mixed} have gained popularity as they allow tasks to have
multiple estimates of execution time at different criticality, or
assurance, levels. Baruah et al.  proposed Adaptive Mixed Criticality
(AMC) as a fixed-priority scheduling policy for mixed-criticality
systems~\cite{baruah2011response}. AMC dominates other fixed-priority
scheduling schemes, such as Static Mixed-criticality with Audsley's
priority assignment and Period Transformation for random
tasksets~\cite{baruah2011response,huang2014implementation}.

AMC scheduling affects the QoS of low-criticality tasks by dropping
them in HI-mode. Further research work explored adjustments to the
task model, including stretching the
periods~\cite{su2013elastic,su2014service,su2016fixed,su2017elastic,gill2018supporting},
and using reduced budget in
HI-mode~\cite{baruah2016scheduling,liu2018scheduling,ramanathan2018multi,baruah2010towards},
to provide improved service to the low-criticality tasks. AdaptMC employs 
control-theoretic feedback to manage the budgets of dual-criticality workloads 
\cite{papadopoulos2018adaptmc}. In contrast, \oursystem{} adjusts the budget of 
a currently running task based on the observed delay at a checkpoint, as we 
want a simple budget adjustment mechanism to minimize the runtime overhead. 
However, future work may explore integrating control-theoretic feedback, such 
as the one used by AdaptMC, into \oursystem{}'s runtime strategy.

Santy et al. proposed the idea of a statically-calculated task
allowance~\cite{santy2012relaxing}, for the theoretical modeling and
scheduling of mixed-criticality tasks. In contrast to Santy's work,
\oursystem{} dynamically decides whether a task is given extra budget
in LO-mode based on the observed runtime delay at a
checkpoint. \oursystem{} is then able to decide which task's budget to
extend in LO-mode, given the slack in computational
resources~\cite{de2009scheduling}.

The work by Kritikakou et al. uses run-time monitoring and control in
mixed-criticality systems, to increase task 
parallelism~\cite{kritikakou2014distributed,kritikakou2014run,
kritikakou2017dynascore}. The authors run high- and low-criticality
tasks together and monitor high-criticality tasks at multiple {\em
observations points} embedded into their control flow graphs.  If
interference from the low-criticality tasks is too prohibitive for the
high-criticality tasks, low-criticality tasks are stopped to ensure
that the high-criticality tasks meet their deadlines.  The authors use
static execution time analysis to decide whether to run the
low-criticality tasks after an observation point in a high-criticality
task. In our work, we dynamically adjust LO-mode budgets of the
high-criticality tasks when we detect delays at intermediate
checkpoints. We decide about the LO-mode budgets based on the observed
progress, instead of using static offline remaining time as used by
other work. In addition, we have implemented an LLVM compiler pass to
\emph{automatically} instrument checkpoints in C and C++ programs, which have 
been tested with a \oursystem{} implementation in \litmusrt{} for real-world 
applications developed for Linux. \annotatered{Kritikakou et al.'s research was 
implemented for a specific DSP platform, which is yet to be tested for 
real-world applications.} Notwithstanding, the authors provide an important 
WCET analysis using CFGs for high-criticality tasks.

%The authors run high- and low-criticality
%tasks together and monitor high-criticality tasks at
%multiple \emph{observations points} embedded into their control flow
%graphs. If interference from the low-criticality tasks is too
%prohibitive for the high-criticality tasks, low-criticality tasks are
%stopped to ensure that the high-criticality tasks meet their
%deadlines. The authors use static execution time analysis to decide
%whether to run the low-criticality tasks after an observation point in
%a high-criticality task.  In our work, we dynamically adjust LO-mode
%budgets of the high-criticality tasks when we detect delays at
%intermediate checkpoints. We decide about the LO-mode budgets based on
%the observed progress, instead of using static offline remaining time
%as used by the other work. In addition, we have implemented an LLVM
%compiler pass to instrument checkpoints in real-world applications,
%which have been tested with a \oursystem{} implementation in 
%\litmusrt{}. Notwithstanding,
%Kritikakou et al. provide an important WCET analysis using CFGs for
%high-criticality tasks.

Previous ideas of progress-based scheduling were proposed to improve
GPU performance~\cite{anantpur2015pro,jeong2012qos}, fairness among
multiple threads~\cite{feliu2015addressing, feliu2017perf} and to
account for instruction cycles~\cite{eyerman2009per}.  Most of these
works run a task in an isolated environment and compare its progress
to an online parallel execution with other tasks. Somewhat similar to
the motivation behind Jeong et al's work~\cite{jeong2012qos}, we also
meet the deadlines of high-criticality tasks.  However, \oursystem{}
uses CFGs to monitor progress rather than application-specific
features such as frame processing rates in multimedia applications as
done in the other works.

\vspace{-4pt}
\section{Conclusions and Future Work}
\label{sec:conclusion}
\vspace{-4pt} This paper presents \oursystem{}, a scheduling strategy
based on the execution progress of a task.  Progress is measured by
observing the time taken for a program to reach a designated
checkpoint in its control flow graph (CFG).  We integrate \oursystem{}
in mixed-criticality systems by extending AMC scheduling.
\oursystem{} extends the LO-mode budget of a high-criticality task
based on its observed progress, given that the extension does not
violate the schedulability of any tasks.  Our extension to AMC
scheduling, called AMC-\oursystem{}, is shown to improve the QoS of
low-criticality tasks.  Moreover, we implement an algorithm in the
LLVM compiler to automatically detect and instrument viable program
checkpoints for use in task profiling.

\annotatered{This paper presents the first implementation of AMC
scheduling in \litmusrt{}. We have also implemented AMC-\oursystem{}
in \litmusrt{} and compared it against AMC.} While both meet
deadlines for all high-criticality tasks, AMC-\oursystem{} improves
the average utilization of low-criticality tasks by 1.5--9 times for
2--20 total tasks.  AMC-\oursystem{} is shown to improve
performance for low-criticality tasks while reducing the number of
mode switches.  Finally, we have shown that different progress metrics
also improve the LC tasks' utilization.

In future work, we will explore other uses of progress-aware
scheduling in timing-critical systems. We plan to extend the Linux
kernel \texttt{SCHED\_DEADLINE}
policy~\cite{sched-deadline,lelli2011efficient,lelli2016deadline} to
support progress-aware scheduling. We believe that \oursystem{} is
applicable to timing-sensitive cloud computing applications, where it
is possible to adjust power (e.g., via Dynamic Voltage Frequency
Scaling) based on progress. Application of \oursystem{} to domains
outside real-time computing will also be considered in future work.

%\section{Additional Algorithms and Evaluations}
%\input{additional.tex}

%\section{Acknowledgments}

%IEEEtran is for sorted, IEEEtranS is normal
\bibliography{references}

%\appendices

\end{document}